\documentclass[useAMS,usenatbib]{mn2e}
\usepackage{epsf}
\usepackage{natbib}
\usepackage{epsfig}
\usepackage{subfigure}
\usepackage{amsfonts}

\def\galform{{\small GALFORM}}
\def\cloudy{{\small CLOUDY}}
\def\subfind{{\small SUBFIND}}
\def\gsim{ \lower .75ex \hbox{$\sim$} \llap{\raise .27ex \hbox{$>$}} }
\def\lsim{ \lower .75ex \hbox{$\sim$} \llap{\raise .27ex \hbox{$<$}} }

\title[{Milky Way satellites in $\Lambda$CDM cosmology}]{The population of Milky Way satellites in the $\Lambda$CDM cosmology} 

\author[A. S.~Font et al.]
{\parbox{\textwidth}{A. S.~Font,$^{1,3}$\thanks{E-mail: \texttt{afont@ast.cam.ac.uk}}
A. J.~Benson$^{2}$,
R. G.~Bower$^{3}$,
C. S.~Frenk$^{3}$,
A.~Cooper$^{3,9}$,
G.~DeLucia$^{4}$,
J. C.~Helly$^{3}$
A.~Helmi$^{5}$,
Y.-S.~Li$^{5,6}$,
I. G.~McCarthy$^{1,7}$,
J. F.~Navarro$^{8}$,
V. Springel$^{10,11}$,
E.~Starkenburg$^{5}$,
J.~Wang$^{3}$,
and
S. D. M.~White$^{9}$}\vspace{0.4cm}\\
\parbox{\textwidth}{$^{1}$Institute of Astronomy, University of Cambridge, Madingley Road, Cambridge, CB3 0HA\\
$^{2}$California Institute of Technology MC 350-17, 1200 E. California Blvd., Pasadena, CA 91125, U.S.A.\\
$^{3}$Institute of Computational Cosmology, Department of Physics, University of Durham, Science Laboratories, South Road, Durham DH1 3LE\\
$^{4}$INAF-Osservatorio Astronomico di Trieste, Via Tiepolo 11, I-34131 Trieste, Italy\\
$^{5}$Kapteyn Astronomical Institute, University of Groningen, PO Box 800, 9700 AV Groningen, the Netherlands\\
$^{6}$Kavli Institute for Astronomy and Astrophysics, Peking University, Beijing 100871, China\\
$^{7}$Kavli Institute for Cosmology, University of Cambridge, Madingley Road, Cambridge, CB3 OHA\\
$^{8}$University of Victoria, 3800 Finnerty Road, Victoria, BC V8P 5C2, Canada \\
$^{9}$Max Planck Institut f\"{u}r Astrophysik Karl-Schwarzschild-Str. 1 85741 Garching, Germany\\
$^{10}$Heidelberg Institute for Theoretical Studies, Schloss-Wolfsbrunnenweg 35, 69118 Heidelberg, Germany\\
 $^{11}$Zentrum f\"ur Astronomie der Universit\"at Heidelberg, Astronomisches
Recheninstitut, M\"{o}nchhofstr. 12-14, 69120 Heidelberg, Germany}}

\begin{document}

\date{Accepted ... Received ...}

\pagerange{\pageref{firstpage}--\pageref{lastpage}} \pubyear{2009}

\maketitle

\label{firstpage}

\begin{abstract}
We present a model for the satellites of the Milky Way in
which galaxy formation is followed using semi-analytic techniques
applied to the six high-resolution N-body simulations of galactic
halos of the Aquarius project. The model, calculated using the
\galform\ code, incorporates improved treatments of the relevant
physics in the $\Lambda$CDM cosmogony, particularly a self-consistent
calculation of reionization by UV photons emitted by the forming
galaxy population, including the progenitors of the central
galaxy. Along the merger tree of each halo, the model calculates gas
cooling (by Compton scattering off cosmic microwave background
photons, molecular hydrogen and atomic processes), gas heating (from
hydrogen photoionization and supernova energy), star formation and
evolution. The evolution of the intergalactic medium is followed
simultaneously with that of the galaxies. Star formation in the more 
massive progenitor subhalos is suppressed primarily by supernova feedback, 
while for smaller subhalos it is suppressed primarily by photoionization 
due to external and internal sources. The model is constrained to match a 
wide range of properties of the present day galaxy population as a whole, 
but at high redshift it requires an escape fraction of UV photons near unity 
in order completely to reionize the universe by redshift $z\gsim 8$. In the 
most successful model the local sources photoionize the pre-galactic region 
completely by $z \simeq 10$. In addition to the luminosity function of Milky 
Way satellites, the model matches their observed luminosity-metallicity
relation, their radial distribution and the inferred values of the mass
within 300~pc, which in the models increase slowly but significantly
with luminosity. There is a large variation in satellite properties
from halo to halo, with the luminosity function, for example, varying
by a factor of $\sim 2$ among the six simulations.

\end{abstract}

\begin{keywords}
{Galaxy: evolution  ---  Galaxy: formation --- galaxies:  dwarf} 
\end{keywords}

\section{Introduction}

A basic prediction of the cold dark matter (CDM) theory of structure
formation is that galactic dark matter halos grow by the accretion and
disruption of smaller subsystems. A well-known consequence of this is
that the Milky Way today should be surrounded by thousands of small
subhalos, in apparent contradiction with the much smaller number of
luminous satellites that have been detected around the Milky Way so
far.  This has been termed the `missing satellite problem'
(\citealt{klypin99,moore99}). A second potential discrepancy with the
theory is the recent observation by \cite{strigari08} that the
satellites of the Milky Way all have very similar central densities
even though they span five orders of magnitude in luminosity, suggesting
the existence of a preferred scale which is not present, for example,
in the primordial cold dark matter power spectrum of density
perturbations.

It has long been recognized that the reionization of the intergalactic
medium (IGM) by the metagalactic UV radiation field at early times can
inhibit the formation of small galaxies
(\citealt{couchman86,efstathiou92,thoul96}). It has
also been recognized for some time that this process would provide at
least part of the solution to the missing satellite problem
(\citealt{kauffmann93,bullock00,benson02b,somerville02}). 
\cite{bullock00}, in particular, emphasized the vital role of early
reionization and calculated a simple model for the abundance of
satellites in the CDM cosmogony.  Since reionization imposes a scale
in the problem---a minimum entropy for the reionized hydrogen---this
process may also solve the second `satellite problem'
(\citealt{li09,okamoto09,stringer10}).

Galaxy formation in small halos can also be strongly inhibited by
winds generated by the injection of supernova energy into the gas
\citep{white78}. The combined effects of this 
so-called ``supernova feedback'' and early reionization were used by
\cite{kauffmann93} to explain 
the relative paucity of satellite galaxies around the Milky Way, and
modelled in greater detail by
\cite{benson02b} who showed that this relatively simple physics could
account not only for the abundance of the `classical satellites', but
also for many of the properties known at the time, such as gas
content, metallicity and star formation rate. The 
\cite{benson02b} model predicted the existence of many more satellites 
in the Local Group. A few years later, a new population of satellites
was discovered, through careful searches of the Sloan Digital Sky
Survey, more than doubling the number of known satellites around the
Milky Way (\citealt{willman05,zucker06,belokurov07,irwin07,walsh07};
see also
\citealt{martin04}).  The luminosity function of the newly found
satellites matches the theoretical predictions remarkably well
\citep{koposov08}.

The discovery of a new population of ultra-faint satellites led to
renewed interest in the physics of dwarf galaxy formation. A number of
recent studies have revisited the arguments of the 1990s and early
2000s, either using semi-analytic modelling techniques
(\citealt{cooper10,guo11a,li10,maccio10}), or simpler models
(\citealt{koposov09,busha10,munoz09}), generally confirming the
conclusions of the earlier work. Some of these studies
(\citealt{busha10,munoz09,cooper10}) have taken advantage of a new
generation of high resolution N-body simulations which track the
formation of galactic halos and their surviving subhalos down to the
smallest halos likely to support the formation of faint galaxies
(\citealt{diemand07,springel08b}). This allows calculation of the
expected radial distribution of satellites which was not predicted
with sufficient precision by the semi-analytic approach of
\cite{benson02b}, based on
Monte-Carlo halo merger trees and an analytic model of satellite
orbital evolution. In addition to these semi-analytic calculations,
full N-body/gasdynamic simulations of galaxies and their satellites
have now been carried out
(\citealt{libeskind07,okamoto09,okamoto10,wade10}). These too find
that reionization plays a key role in suppressing dwarf galaxy
formation but, as stressed by \cite{kauffmann93}, feedback from
supernova energy is also a major factor. \cite{wade10} find, further,
that cosmic ray pressure is also important in suppressing star
formation in these galaxies.

Although a consensus seems to be emerging then, that the abundance and
other properties of the Milky Way satellites can be understood as a
consequence of the known physics of galaxy formation in a $\Lambda$CDM
universe, a number of important uncertainties remain. For
example, \cite{boylan-kolchin11} have recently concluded, on the basis
of dynamical data, that the most massive subhalos in N-body
simulations are too concentrated to be able to host the brightest
satellites of the Milky Way. There are also significant modelling
uncertainties. Recent studies treat reionization in a
simplified way
(\citealt{busha10,koposov09,maccio10,busha10,guo11a,li10}), reducing
this complex process to a simple rule involving a few adjustable
parameters (e.g., a single redshift for reionization and a threshold
mass above which reionization is unimportant). Furthermore, the
luminosity function is just one of many observables that depend
critically on the assumed physics of the problem. The metallicity of
the satellites, for example, not considered in most of these studies,
is very sensitive to the effects of supernova feedback.

In this paper, we investigate the formation of satellite galaxies in
the $\Lambda$CDM cosmogony employing a new treatment of reionization,
based on that of \citet{benson02a}, that simultaneously solves for the
coupled evolution of the galaxy population and the IGM. The ionizing
background is computed from the integrated emission history of all
galaxies in the Universe plus a simple model for the quasar
contribution, filtered through the optical depth of the IGM whose
evolution and ionization state are computed self-consistently. The
ionizing radiation, generated by galaxies and quasars, feeds back upon
subsequent galaxy formation through the dual action of
photoionization, which {\it i)}~prevents the collapse of gas onto low
mass dark matter halos and {\it ii)} reduces the rate of radiative
cooling inside more massive halos. An original feature of this model
is the inclusion of internal reionization from progenitors of the
Milky Way.

Since the early work on satellite galaxies a decade ago, there have
been important developments in techniques for modelling galaxy formation. Many
of these have been incorporated into the Durham semi-analytic model,
\galform\ (\citealt{cole00,bower06}), which we use in this paper. To this, we add 
treatments of cooling and heating processes that are relevant on the
scale of dwarf galaxies. In addition to the usual atomic radiative
processes, we include cooling due to molecular hydrogen ($H_2$) and
Compton cooling of gas due to scattering off cosmic microwave
background photons.  Heating due to photoionization is calculated
using \cloudy\ \citep{ferland98}. The evolution of the IGM is computed
taking into account the contribution of the UV ionizing background
from both galaxies and quasars. The emissivity from galaxies is
calculated self-consistently, whereas the contribution from quasars is
modelled using an updated version of the spectrum inferred
observationally by
\citet{haardt96}. The model then predicts the time dependence of
the properties of the IGM, such as its mean temperature and ionization
state which, in turn, determine the rate at which it cools into galaxies.

We find that, in addition to the global UV background pervading the 
Universe, the radiation generated locally by stars forming
in the progenitors of the Milky Way is an important source of ionization 
in the Milky Way at early times.  As we will
demonstrate, this radiation increases the redshift of
reionization locally and leads to a stronger suppression of satellite
galaxy formation than the metagalactic flux.  Recently,
\citet{munoz09} have identified this mechanism of `inside out
reionization' of the Milky Way as important in suppressing the
formation of ultra-faint dwarfs.  According to these authors,
self-reionization occurs much before the time when the Universe as a
whole was reionized, at an epoch when most of the gas was in molecular
($H_2$) form. They suggest that reionization would quench cooling of
the $H_2$ gas, stopping it from fragmenting and forming stars,
possibly explaining the paucity of ultra-faint dwarfs today. In
practice, they model this process by making similar assumptions to 
those made in simple models of the effects of global reionization,
namely by assuming that $H_2$ cooling becomes ineffective below a
given halo mass after a given redshift. 
By contrast, we implement the self-reionization of the Milky Way by 
explicitly calculating the UV flux produced by the progenitors of the 
Milky Way as a function of time.  This takes away any freedom in the choice 
of redshift and halo mass for which this process is important, while  
the same is achieved concomitantly for the global reionization with the 
self-consistent approach.

\begin{table*}
\begin{tabular}{l | c | c | c | c | c}
\hline
Halo & $m_p$ [M$_{\odot}$] & $N_{hr}$ & $M_{200} [M_{\odot}]$ &
$r_{200}$ [kpc] & $M_{*,tot}[M_{\odot}]$ \\ 
name & & & & & fbk:sat/rei:G+L ($f_{\rm esc}=1$)\\
\hline		
  AqA & $1.370 \times 10^4$ & 531,570,000 & $1.842 \times 10^{12}$ & 245.88 & $1.085 \times 10^{11}$ \\
  AqB & $6.447 \times 10^3$ & 658,815,010 & $8.194 \times 10^{11}$ & 187.70 & $ 9.179 \times 10^{9}$ \\
  AqC & $1.399 \times 10^4$ & 612,602,795 & $1.774 \times 10^{12}$ & 242.82 & $ 1.050 \times 10^{11}$ \\
  AqD & $1.397 \times 10^4$ & 391,881,102 & $1.774 \times 10^{12}$ & 242.85 & $ 6.982 \times 10^{10}$  \\
  AqE & $9.593 \times 10^3$ & 465,905,916 & $1.185 \times 10^{12}$ & 212.28  & $ 2.217 \times 10^{10}$ \\
  AqF & $6.776 \times 10^3$ & 414,336,000 & $1.135 \times 10^{12}$ & 209.21  & $ 6.555 \times 10^{9}$ \\
\hline  
\end{tabular}
\caption{\label{table:Aq}{Properties of the Aquarius halos, as given 
by \citet{springel08b}. Listed here are: the particle mass, $m_p$; the
total number of particles in the high resolution region, $N_{hr}$; the
virial mass of the halo, $M_{200}$; and its virial radius,
$r_{200}$. The last column gives the total stellar mass for each
Aquarius halo in the fiducial model, fbk:sat/rei:G+L ($f_{\rm esc}=1$), 
that will be described later in the paper. 
The stellar masses in this model show a large scatter around the value
estimated for the Milky Way, $4.85 - 5.5 \times 10^{10} M_{\odot}$ by
\citet{flynn06}.}}
\end{table*}  

In this paper, we implement our model of the coupled evolution of
galaxies and the IGM in merger trees constructed from the high
resolution simulations of galactic halos in the Aquarius project
\citep{springel08a}. The resolution of these simulations 
(with particle masses ranging from $6.4\times
10^{3}$ to $1.4\times 10^{4} M_{\odot}$) is sufficient
to enable predictions to be made for satellites as faint as even the
faintest dwarfs detected recently around the Milky Way. Using the six
Aquarius simulations, we can investigate halo-to-halo variations in
the properties of the satellites. 

Our method for simulating the joint evolution of galaxies and the IGM,
as well as the new physics we have implemented in \galform, are
explained in Section~\ref{sec:theor}. In Section~\ref{sec:results}, we
demonstrate how the metallicity$-$luminosity relation of satellites
can be used to break the model degeneracy between the effects of
supernova feedback and reionization. We also explore how
self-reionization changes the predicted abundance of satellites. We
find, however, that it is not possible for the model simultaneously to
match the luminosity function and the metallicity$-$luminosity
relation with the default supernova feedback prescription used in
previous implementations of
\galform, in which the efficiency of the feedback increases with
decreasing halo size as a power-law. Instead, the satellite data
require that this feedback efficiency should saturate for halos with
circular velocity, $v_{\rm circ} \leq 65$ km s$^{-1}$. In Section~4, we
test this model further by considering predictions for the radial
distribution of satellites and for the mass within 300~pc. Finally, in
Section~5, we discuss our conclusions and carry out a comparison with
the results by
\cite{li10} using similar semi-analytic techniques. A detailed
comparison with other related models is presented in Appendix~B.

The cosmological parameters adopted in this study are the same as
those used for the Aquarius and Millennium \citep{springel05b}
simulations and correspond to a flat $\Lambda$CDM cosmology with
matter density, $\Omega_{m}=0.25$, cosmological constant term,
$\Omega_{\Lambda}=0.75$, Hubble parameter $h=H_{0}/$ (100 km s$^{-1}$
Mpc$^{-1}$) = 0.73, power spectrum normalization, $\sigma_{8}=0.9$,
and spectral index $n_s=1$\footnote{These parameters were chosen
to match the WMAP 1st-year results. Of these, only $\sigma_{8}$ and
$n_s$ are of importance for our study. The adopted value of
$\sigma_{8}$ leads to an earlier redshift of reionization than
expected for the WMAP 7-year value ($\sigma_{8}=0.80$), but this is
partly compensated for by the larger value assumed for $n_s$ (0.96 from
the WMAP 7-year data). In any case, these differences are expected to
be small \citep{wang08,boylan-kolchin10,iliev11}, especially compared with
those arising from uncertainties in the modelling of reionization 
(e.g., the assumed escape fraction or the clumping factor).}.

\section{Theoretical models}
\label{sec:theor}

We begin by describing the Aquarius N-body simulations used to build
the merger trees employed by the semi-analytic galaxy
formation model, \galform. We then describe briefly the updates that
we have made to \galform\ itself to include physical processes that
are relevant on the scale of low mass galaxies.

\subsection{The Aquarius halos}
\label{sec:Aqhalos}

In the Aquarius project, carried out by the Virgo Consortium, six
galactic dark matter halos with masses comparable to that of the Milky
Way were simulated at varying levels of resolution
\citep{springel08a,springel08b}.  Here, we use all six halos at the
second highest level of resolution (`Level 2' in
\citealt{springel08b})\footnote{The highest level of resolution, `Level
1', is available only for one of the halos.}. 
We define as `satellites' all sub-halos found by
\subfind\, \citep{springel01} with more than 20 bound particles. 
Merger trees were constructed for each subhalo using the methods laid
out in \citet{helly03} and \citet{harker06}.

Table \ref{table:Aq} provides a summary of the relevant properties of
the Aquarius halos, taken from \citet{springel08b}. The halos all have
virial masses in the range 1--$2\times10^{12}M_\odot$. Recent evidence
from the analysis of \cite{guo11a} suggests that the current best
estimate of the Milky Way's halo mass is $2 \times 10^{12} M_\odot$,
with a 10\% to 90\% confidence interval spanning the range $0.8$ to
$4.7 \times 10^{12}M_\odot$; other studies, however, favour a mass
closer to $10^{12} M_{\odot}$ \citep{battaglia06a,smith07,xue08}.
Thus, the Aquarius halos sample the lower end of the currently
plausible range of Milky Way halo masses and could underestimate the
true mass by a factor of 2--3. This would have important consequences
for the properties of subhalos and, therefore, for any satellite
galaxies they may contain. We have checked whether the central
galaxies in our models have stellar masses comparable to those
estimated for the Milky Way, $4.85 - 5.5 \times 10^{10} M_{\odot}$
\citep{flynn06}. The final column of Table~\ref{table:Aq} shows
the new fiducial model that will be described later
in more detail. The total stellar masses seem realistic,
although there is significant variation from halo to halo.

\subsection{\galform: The basic model}
\label{sec:basic}

Our starting point is the implementation of 
\galform\ by \citet{bower06}. 
\galform\ includes the treatments, introduced by
\citet{cole00}, of shock-heating, radiative cooling of gas within dark
matter halos (modulo the reduction in cooling rates due to
photoionization described below), star formation, spheroid formation
(through both disk instabilities and galaxy merging), chemical
evolution and dust extinction. In addition, this version contains
prescriptions introduced by \citet{bower06} for feedback from active
galactic nuclei (AGN) and supernovae explosions, and mass loss in
stellar winds. The \citet{bower06} galaxy formation model reproduces
properties of the galaxy population over a wide range of scales and
epochs. For example, it gives a good match to the global luminosity
and stellar mass functions and to the bimodal colour distribution of
galaxies observed in the SDSS and can also account for the redshift
evolution of these properties.

\citet{font08} have found that the relative
distribution of SDSS galaxy colours amongst the blue and red
sequences, as well as the zero-point of the bimodal colour
distribution, can be better matched by \galform\ by setting the
metallicity yield to $p=0.04$ (i.e., twice the value adopted by
\citet{bower06}). In addition, increasing the yield by this factor
results in a better match to the metallicity of the intracluster
medium \citep{bower08} and improves the predicted metallicities of the
dwarf galaxies that are of interest here. For these reasons, we adopt
this higher value of the yield in this paper.  As in the study of
\citet{bower06}, we assume that the hot gaseous halos of satellites
are completely and instantaneously stripped by ram pressure as soon as
they cross the virial radius of their host halo. (\citet{mccarthy08}
have shown that ram pressure stripping of hot gas around satellites in
groups and clusters can occur on a relatively long timescale but for
satellites of galaxies like the Milky Way, the assumption of
instantaneous stripping is a good approximation.)

Feedback from supernovae plays an important role in establishing the
properties of Milky Way satellites. In this paper, we investigate
different feedback schemes and how they affect both the luminosity
function and the luminosity - metallicity relation. The first feedback
scheme we analyse is the implementation in
\galform\, presented in \citet{bower06} (hereafter the ``default
model'' or fbk:B06). In the \citet{bower06} model, supernovae are
assumed to inject energy into the cold gas in the disk of the galaxy,
heating it to the virial temperature of the halo after which the gas
is ejected. This `reheated' gas may subsequently cool and re-settle
into the disk. The efficiency of stellar feedback, $\beta$, is assumed
to depend on the circular velocity of the halo, $v_H$, in which the
galaxy resides as $\beta=(v_{H}/v_{\rm hot})^{-\alpha_{\rm hot}}$ (see
also
\citealt{cole00}). The parameters $v_{\rm hot}$ and $\alpha_{\rm hot}$
control the feedback efficiency, and in the \citet{bower06} model,
they are set to $v_{\rm hot}=485$ km s$^{-1}$ and $\alpha_{\rm
hot}=3.2$.  Note that larger values of $\alpha_{\rm hot}$ correspond
to a greater effectiveness of supernova feedback in halos with $v_{H}
< v_{\rm hot}$, which is the regime of interest here.

As we shall see in Section~\ref{sec:sf}, the stellar feedback assumed in 
\citet{bower06} leads to model satellites with metallicities lower than 
observed. This happens because the supernova feedback in this model is 
too efficient at expelling metals from dwarf galaxies. We find that 
simply decreasing the slope of the $\beta(v_{H})$ dependence 
is not a viable solution to this problem: while reasonable matches to 
the satellite metallicities can be found with, for example, 
$\alpha_{\rm hot} \simeq 2.5$, this approach leads to a boost in the number of 
field dwarf galaxies which is inconsistent with data from large-scale surveys.
 
\begin{figure}
\includegraphics[width=9cm]{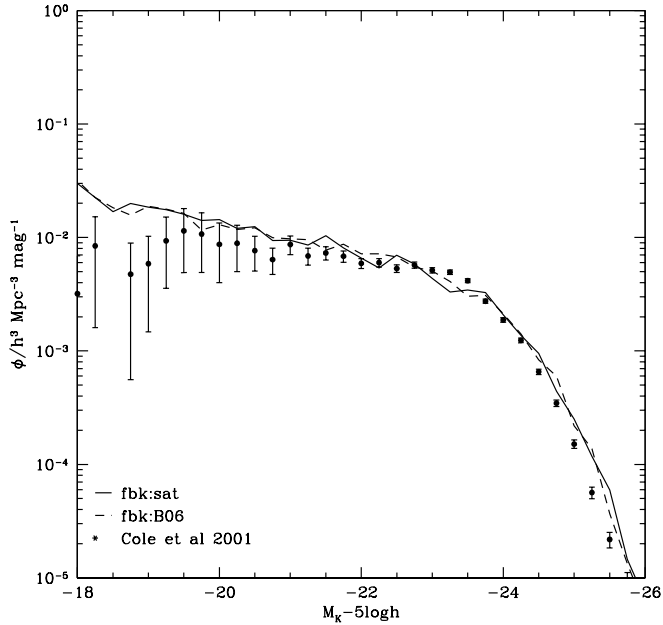}
\caption{\label{fig:global_KLF}{Comparison of the global K-band luminosity 
function for the fiducial model with saturated feedback used in this paper 
(full line) and the \citet{bower06} model (dashed line). 
The models are compared with observational data from \citet{cole01}.
Although the two models differ in their treatment of stellar feedback below 
$v_{\rm circ} =65$ km s$^{-1}$, they provide entirely comparable fits to 
the observational data at brighter magnitudes.}}
\end{figure}

We consider the possibility that supernova feedback may not scale as a
simple power law across the full range of galaxy masses and
investigate alternative formulations that preserve the good agreement
with large-scale data that was the trademark of the \citet{bower06}
model. In Section~\ref{sec:fbksat} we find a viable model which has
the property that the efficiency of stellar feedback, $\beta$,
saturates in small mass halos, with $v_{\rm circ} \leq 65$ km
s$^{-1}$, while above this value it retains the behaviour of the
default  model. This model is called fbk:sat. It
matches the global properties of the galaxy population well (see
Fig.~\ref{fig:global_KLF}, for example) and, as we will show below,
also a variety of Local Group data, including the flattening of the
metallicity-luminosity relation observed for the ultra-faint
dwarfs. The stellar feedback is required to be very inefficient in
these systems. 

It seems plausible that the feedback efficiency is low in small
mass galaxies because their star formation rate (and hence the
supernova rate) is observed to be low. Dwarf galaxies may be
pre-enriched by Population~III stars to a level of order
$10^{-3}Z_\odot$ \citep{wise10}.  Afterwards, due to the
inefficient feedback, the smallest dwarfs would evolve
approximately as a closed-box system
\citep{salvadori09}, and a single generation of supernovae may be sufficient
to increase their metallicity by $\sim 0.5$~dex, as observed.  We note
that a similar saturation scheme has been proposed recently by
\cite{guo11a}, who implemented it in the Munich semi-analytic 
model. Theirs is the first model that successfully matches both the
large-scale data and the properties of dwarf galaxies, although only
at the resolution of the Millennium-II simulation, $m_p
\simeq 9.4\times 10^6M_\odot$ \citep{boylan-kolchin09}.
Table~\ref{table:Models} summarizes the main feedback and reionization
features of the models used in the paper.

\begin{table*}
\begin{tabular}{l | c| c | c | c}
\hline
Model & Supernova feedback & Global & Local & Escape \\
name & efficiency ($\beta$) & reionization & reionization & fraction ($f_{\rm esc}$)\\
\hline		
  fbk:B06/rei:G & \citet{bower06} & $\checkmark$ & $\times$ & 1 \\
  fbk:B06/rei:G+L & \citet{bower06} & $\checkmark$ & $\checkmark$ & 1 \\
  fbk:sat/rei:G & Saturated & $\checkmark$ & $\times$ & 0.1, 0.2, 0.5, 1 \\
  fbk:sat/rei:G+L & Saturated & $\checkmark$ & $\checkmark$ & 0.1, 0.2, 0.5, 1  \\
\hline
\end{tabular}
\caption{\label{table:Models}{Features of the models. The name of
each model reflects whether it includes the default supernova
feedback (fbk:B06) efficiency of \citet{bower06} or the saturated feedback
efficiency (fbk:sat), whether it includes global (rei:G) or local (rei:L)
reionization. (Details of the reionization model are given in
\ref{sec:add_phys}.)}}
\end{table*}  

When modelling galaxy formation in lower resolution simulations than
Aquarius care must be taken to follow satellite galaxies in subhalos
that are disrupted below the adopted particle limit (20 in our case)
by tidal forces. In the terminology of \citet{springel01}, these
galaxies are of `type 2' while galaxies present in identified dark
matter subhalos are of `type 1'. At the resolution of Aquarius ($\sim
10^4 M_\odot$), we can safely assume that whenever a subhalo is
disrupted so is the galaxy (if any) that it contains, that is, there
are no type-2 galaxies. Changes in the structure of subhalos caused by
baryons cooling inside them are likely to be negligible since
satellites have very large mass-to-light ratios. We have checked that
the sizes of potential type-2 satellites given by \galform\,
(including the effects of adiabatic contraction caused by the bayrons)
immediately before their subhalos are disrupted are generally larger
than the subhalo tidal radius. We therefore assume that `type 2'
galaxies are disrupted when their halos are disrupted and exclude them
from further analysis.

With the exception of the additional physical processes discussed
below, all the other parameters in our semi-analytic model are the
same as in \citet{bower06}.

\subsection{Additional physics}
\label{sec:add_phys}

Here, we briefly describe the inclusion of new physical processes,
relevant on dwarf galaxies scales, in the \galform\ model.

\subsubsection{Cooling, reionization, and photoheating} 

The cooling function, $\Lambda(\rho,T,Z,z)$ (which is the net cooling
rate of gas, obtained by summing the cooling and heating terms), is
calculated self-consistently at each redshift using the photoheating
background predicted by our model of the evolution of the
intergalactic medium (IGM) as described below. To carry out this
calculation we employ the code \cloudy\ \citep{ferland98}. The
radiative cooling processes included are: Compton cooling off the
cosmic microwave background, thermal bremsstrahlung, and the usual
atomic radiative processes. We also include molecular ($H_2$) cooling
following the prescription of \citet{benson06}.

Our model for the evolution of the IGM is essentially that described
in \citet{benson06} (see also \citealt{benson10}), which follows the
photoionization and recombination of the IGM in addition to cooling
and heating rates thereby allowing the ionization and thermal state of
the IGM to be predicted as a function of time. The photoionizing flux
present at any point within an Aquarius halo is made up of two
contributions, a local one due to sources within the halo and a global
one due to sources in the rest of the Universe. The calculation of the
local flux is explained in Section~\ref{sec:local}. To obtain the
global flux, we run the \galform\ code on a very large set of
Monte-Carlo merger trees generated using the empirical modification of
the extended
\citet{press-schechter} formalism advocated by \citet{parkinson08}, 
which gives results consistent with N-body simulations. This method
gives the properties of the IGM outside the Aquarius halo, as well as
the global photoheating background. At each timestep in the \galform\
calculation, we determine the mean emissivity (as a function of
wavelength) per unit volume in the Universe by summing the
contributions of galaxies and quasars. This emissivity is used to
compute the rate at which the background of ionizing photons is built
up. The background can experience absorption by neutral gas in the
IGM, and so it is strongly suppressed prior to reionization.

For galaxies, the emissivity is obtained self-consistently from the
population of galaxies formed by \galform\ up to the current
timestep. The efficiency with which this radiation ionizes the IGM
depends on the assumed escape fraction of ionizing photons. While both
observational and theoretical estimates of this fraction exist
(e.g. \citealt{atek09,laursen09,siana07,wise09}), its value remains
highly uncertain, particularly at high redshift. We therefore chose to
treat the escape fraction as a free parameter which we fix so as to
produce a reionization history compatible with experimental
constraints \citep{komatsu10}. As we will show in
Section~\ref{sec:fbksat}, we find that our new model requires a high escape
fraction ($ \sim 80 - 100\%$) in order to match the present number of
satellite galaxies. The high escape fraction reflects the fact that our
model produces too few ionizing photons at high redshifts to reionize
the Universe sufficiently early.

The escape fraction in our model is clearly unrealistically large.
Observational estimates suggest values of $\lsim 10-20$\% 
(e.g. \citealt{siana07,vanzella10}; but see \citealt{bolton07}), but  
we are forced to adopt a value of $\sim$100\% to achieve reionization
by a sufficiently high redshift. While it is unclear whether the observed
values are representative of the galaxy population as a whole (e.g. if
unobserved, low luminosity galaxies dominate the production of ionizing
photons, and their escape fractions are much larger, then the ionizing
emissivity-weighted escape fraction may be much larger), this is nevertheless
a point of tension between our model and current empirical
expectations.

Other models of the formation of the Local Group satellites have not
attempted to model the formation of galaxies and the process of
reionization self-consistently as we have here, and so have been able
either to choose an epoch of reionization consistent with constraints,
or to put in an ionizing emissivity by hand that is designed to give a
reasonable epoch of reionization.

For the purposes of this work, which is concerned primarily with the
Local Group satellite galaxies, and not with the process of
reionization itself, we take the pragmatic view that we should do
whatever is necessary to obtain a realistic epoch of reionization
since it is this that primarily impacts the formation and suppression
of dwarf galaxies. The fact that this requires a large escape fraction
is interesting, and deserves further attention, but is not surprising
given that both observations and our model are highly uncertain in the
high redshift regime. A better model will most likely need to produce more
ionizing photons at high redshift thereby allowing the escape fraction
to be reduced. Of course, a self-consistent calculation of the escape
fraction for model galaxies would provide a more rigorous way to
approach this problem (e.g. \citealt{fernandez11}). A more detailed
study of reionization in the \galform\ model is currently underway
\citep[see][]{raicevic10} and will explore this issue in greater
depth.

For calculating the emissivity of quasars, we use the code CUBA 
\citep{haardt01}, which includes an updated version of the observationally 
inferred \citet{haardt96} spectrum. This approach allows us to track the 
evolution of the IGM while simultaneously calculating the heating rate of the 
gas inside galaxies.

\begin{figure*}
\includegraphics[width=\columnwidth]{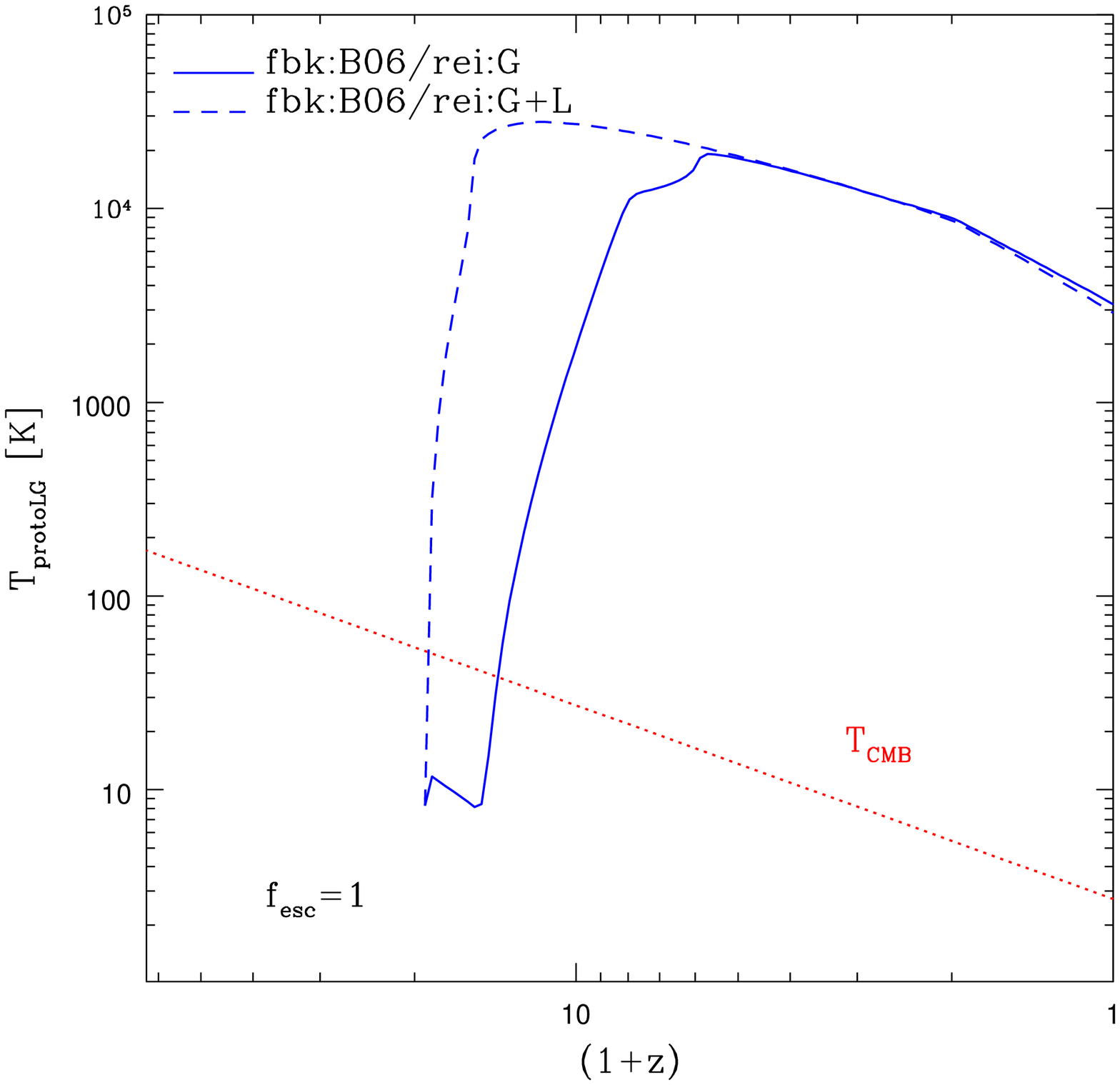} 
\includegraphics[width=\columnwidth]{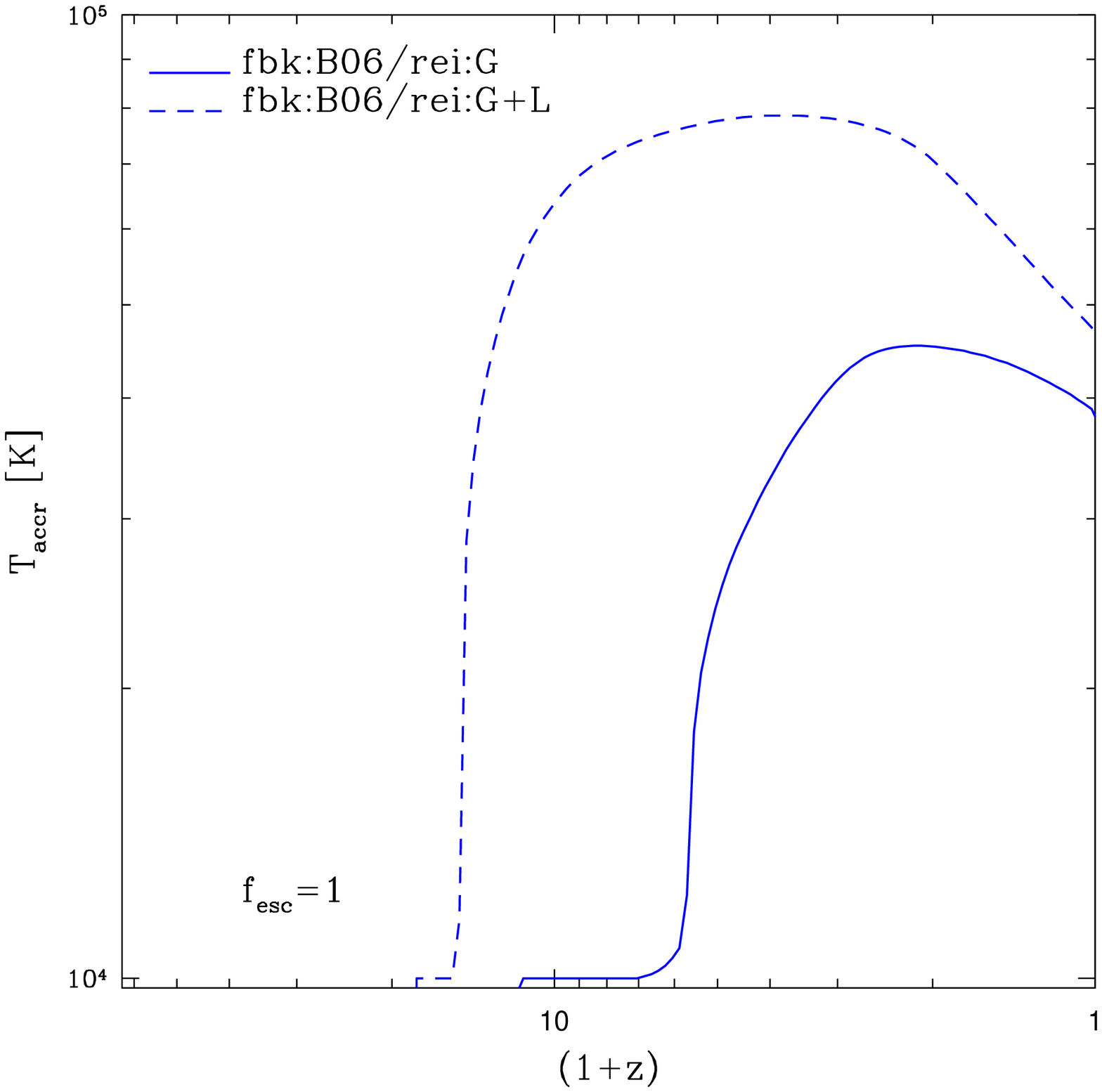}
\caption{\label{fig:IGM_B06}{Evolution of the IGM in the 
default feedback \citep{bower06} model. {\it Left}: the mean
temperature of the gas in the proto-Local Group region in the models
with and without the added local UV flux, as a function of redshift
(dashed and solid blue curves respectively). Reionization occurs at
$z\simeq 14$ in the model with local and global UV flux and at
$z\simeq 6$ in the model with only global UV flux. In the first model,
the temperature peaks at just below $3 \times 10^4$K whereas in the
second model it peaks at $2
\times 10^4$K. Both models assume an escape fraction of 100\%. 
The temperature of the CMB is shown with a red dotted 
line for reference.  {\it Right:} the characteristic accretion 
temperature of gas, $T_{\rm accr}$, as a function of redshift. The 
accretion temperature peaks at $z\simeq 3$, at a value of $8\times 
10^4$K in the model with local UV production, and only at $z\simeq 
1.5$, at a value of $4.5 \times 10^4$K, in the model with only global 
UV production. Note the different scales on the $y$-axis of each 
panel.}}
\end{figure*}

\begin{figure*}
\includegraphics[width=8.cm]{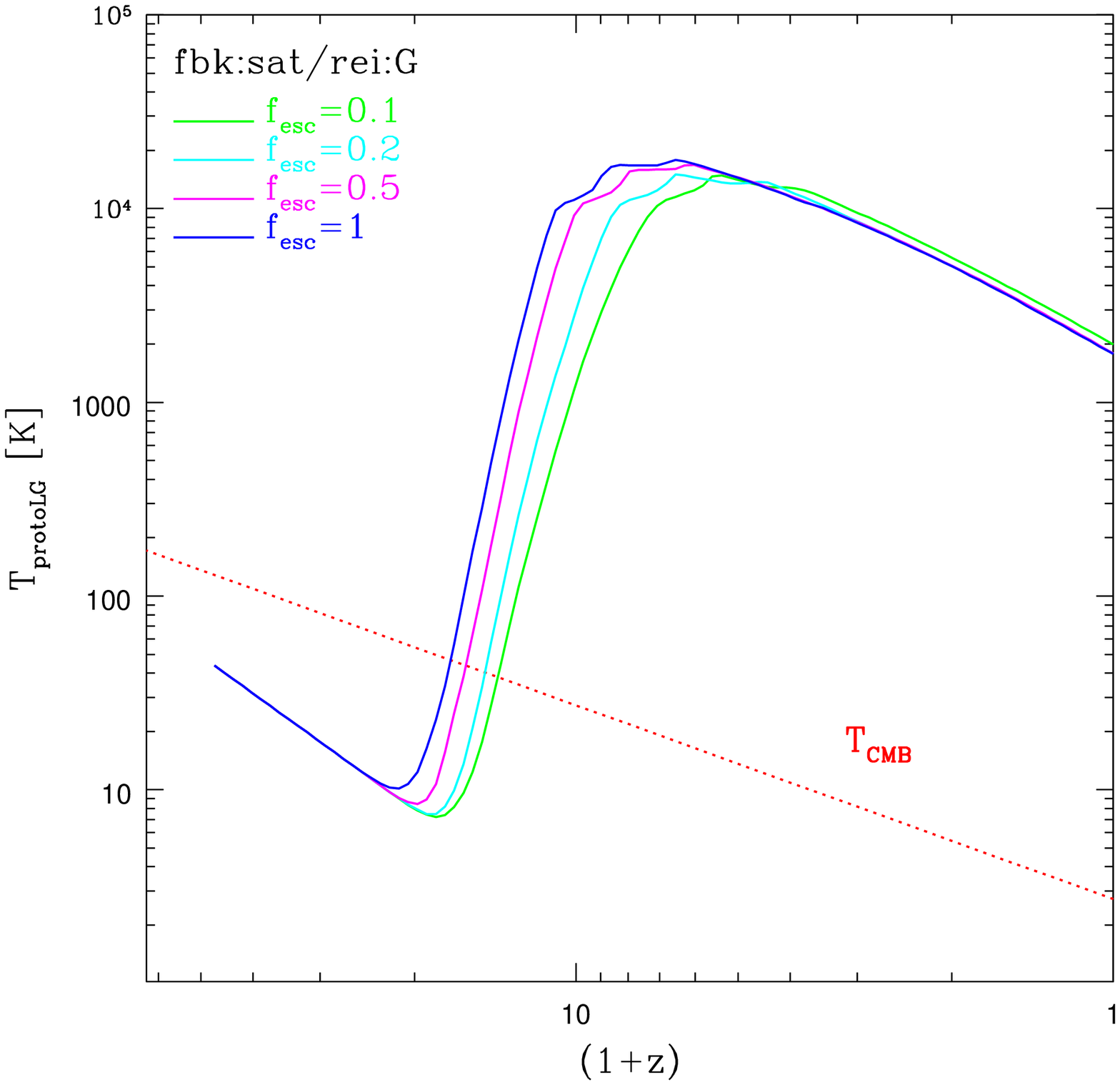} 
\includegraphics[width=8.cm]{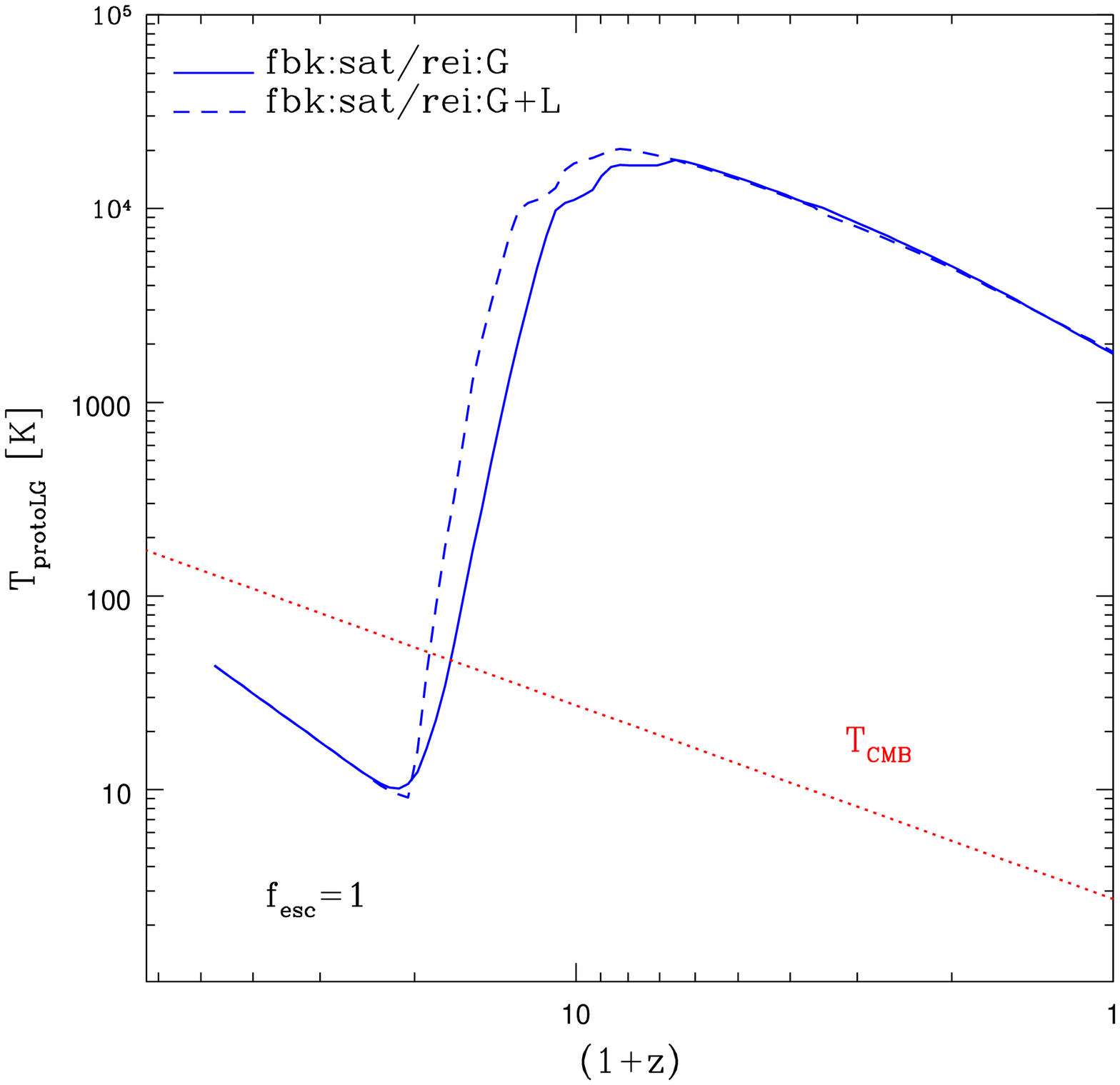}
\caption{\label{fig:IGM_fbksat}{Evolution of the mean temperature of
the gas in the proto-Local Group region in the model with saturated
feedback. The left panel corresponds to the case of global
reionization only, with different colours indicating the escape
fraction. Global reionization occurs at $z\simeq 7.8$ for $f_{\rm
esc}=1$, $z\simeq 6$ for $f_{\rm esc}=0.5$, $z\simeq 5.5$ for $f_{\rm
esc}=0.2$ and $z\simeq 4.5$ for $f_{\rm esc}=0.1$
respectively. $T_{protoLG}$ peaks at $\sim 2
\times 10^4$K for $f_{\rm esc}=1$ and decreases slightly to $\sim 1.5 
\times 10^4$K for $f_{\rm esc}=0.1$. The temperature of the CMB is 
shown with a red dotted line for reference. The right panel shows a
comparison of $T_{protoLG}$ in the saturated feedback model with and
without local reionization, for $f_{\rm esc}=1$ (dashed and solid blue
lines, respectively). Local emission shifts the redshift of
reionization from $z\simeq 7.8$ to $z\simeq 10$, without a significant
increase in the peak of the gas temperature.  }}
\end{figure*}

Heating due to UV photoionization (photoheating, for short) not only
offsets cooling losses but it also prevents low mass dark matter halos
from accreting their full complement of baryons. The suppression of
baryonic accretion into halos is modelled using the accretion mass
scale recently derived by \citet{okamoto08} (see also \citealt{hoeft06})
from cosmological hydrodynamical simulations. These authors find that
the baryon fraction, $f_{\rm b}\equiv M_{\rm b}/M$, depends on halo
mass and redshift as:
\begin{equation}
f_{\rm b}(M,z) = \langle f_{\rm b}\rangle
\left[1+(2^{{\alpha}/3}-1)\left(\frac{M}{M_{\rm
c}(z)}\right)^{-\alpha}\right]^{-3/\alpha}, 
\end{equation}

\noindent where $\langle f_{\rm b}\rangle = \Omega_{\rm b}/\Omega_0$
is the universal baryon fraction, $M_{\rm c}(z)$ is a characteristic
mass that will be defined below, and $\alpha=2$ provides a good fit to the 
results of the simulations. It is worth noting that the \citet{okamoto08} 
accretion mass scale is significantly lower than the ``filtering mass''
calculated previously by \citet{gnedin00} on the basis of linear
perturbation theory. With this new method, halos that accrete only
half of the universal baryon fraction (i.e. $f_{\rm b} = \langle
f_{\rm b}\rangle/2$) have, on average, circular velocities of about 25
km s$^{-1}$ at redshift $z=0$, compared with $50$ km s$^{-1}$ found by
\citet{gnedin00}. In other words, the \citet{okamoto08} mass accretion
scale allows more satellites to escape the effects of the reionization
than the associated filtering mass in the \citet{gnedin00} formalism.
(The latter was used in the previous versions of \galform).  As we
will show, this has important consequences for the inferred role of
supernova feedback on relatively massive satellites.

We follow the method suggested by \citet{okamoto08} for implementing
their scheme of baryon accretion, which involves computing the
equilibrium temperature of gas as it accretes into the halo, $T_{\rm
accr}$. The simple expectation is that the characteristic mass scale
should be set such that $T_{\rm vir}(M_{\rm c}[z],z)=T_{\rm accr}(z)$
where $T_{\rm vir}(M)$ is the virial temperature of a halo of mass $M$
at redshift $z$. However, \citet{okamoto08} show that this approach
overestimates the suppression mass scale and does not accurately
reproduce the redshift dependence found in their simulations. Instead,
they recommend a simple model in which each halo accretes gas at the
universal rate (i.e. $\Omega_{\rm b}/\Omega_0$ times its total mass
accretion rate) if $T_{\rm vir} > T_{\rm accr}$, and accretes no gas
if $T_{\rm vir} < T_{\rm accr}$. We adopt their model in this
work. The accretion temperature is computed self-consistently using
the cooling function described above and the known density of
accreting material. Note that the \citet{okamoto08} formalism provides
the accretion mass scale only after the epoch of reionization, when
there is a significant ionizing background, at which point the authors
suggest setting the critical temperature to the equilibrium
temperature for the photoionized accreting gas. At earlier times,
prior to reionization, we assume that the pre-shock temperature of gas
as it is accreted onto a halo is given by the temperature to which it
is adiabatically heated up from the mean IGM temperature, or $10^4$K,
whichever is lower. This is a valid approximation prior to
reionization as the IGM gas will be mostly neutral, atomic and
metal-free and so there are no efficient cooling processes below
$10^4$K, and little photoheating due to the lack of a significant
ionizing background. At later times, the full complement of heating
and cooling processes are used to compute the accretion temperature.

Our treatment of reionization can be compared to other
semi-analytic/semi-numerical approaches. In particular,
\citet{mesinger11} recently described a semi-numerical algorithm to
compute the ionization state of the IGM. Unlike us,
\citet{mesinger11} compute a position-dependent ionization fraction by
recourse to cosmological simulations which they use to estimate
densities, collapsed fractions and star formation rates (assuming a
constant efficiency to convert collapsed fraction to star
formation). In contrast, our model assumes a uniform reionization, but
additionally solves for the thermal state of the IGM and has a much
more detailed treatment of the underlying galaxy formation physics to
predict the ionizing emissivity. Further advances in modelling of
reionization must combine aspects of these and other approaches. For
example, \citet{raicevic10} compute position-dependent reionization by
combining a detailed model of galaxy formation with ray-tracing
through a cosmological simulation of the density field.

\subsubsection{The local UV flux from Milky Way sources}
\label{sec:local} 

In addition to the global photoionizing flux originating from large
scales (from AGN and quasars), {\it local} sources can also be
important contributors, especially at high redshift.  If these local
sources are significant, reionization in the region will occur earlier
than in an average region of the Universe. Thus, the entire region
destined to become the Local Group may be encompassed within an
ionized bubble (``local reionization''), before many such bubbles have
percolated to reionize the Universe as a whole (``global
reionization''). One consequence of local reionization is to 
suppress further the formation of dwarf satellites within the region.

A fully self-consistent treatment of local reionization requires not
only accounting for the emissivity of local sources, but also
knowledge of other (at the moment, poorly constrained) parameters,
such as the photon escape fraction and the gas clumping factor, and
modelling of radiative transfer.  A comprehensive treatment of this
kind has not been performed so far.  Current models that do include
radiative transfer (e.g. \citealt{weinmann07,iliev11}) have very
simplified physical prescriptions for star formation and feedback.
Semi-analytical models such as \galform, which have more realistic
star formation and feedback prescriptions and which match a wide range
of observations, by contrast, do not include radiative transfer. The
combination of the two approaches is clearly desirable for a complete
understanding of the role of local reionization (see
\citealt{raicevic10} for an example of progress in this area).  Below
we present a simple method for accounting for the local
photons. Without radiative transfer, this is likely to overestimate
the contribution of local photons to the suppression of low-mass
satellites. Nevertheless, we present the results here as proof of
concept of how the redshift of reionization is increased by
contribution from pre-galactic sources.

To calculate the contribution to the ionizing flux from local sources,
we identify the galaxy's progenitors by finding halos in the merger
tree that lie within the virial radius of the final Aquarius halo at
$z=0$. We then calculate the total luminosity of ionizing photons
emitted by these local sources at all redshifts, and compute the
effective photon density within the Lagrangian radius of the final
halo:

\begin{equation}
n (z) = \frac{L_{\rm ion} (z)}{4 \pi {R^{2}}_{\rm Lag}(z) {\rm c}} \, ,
\end{equation}

\noindent where $L_{\rm ion}(z)$ is the total instantaneous ionizing luminosity, ${\rm
c}$ is the speed of light and $R_{\rm Lag}(z)$ is the radius of a
sphere with volume equal to the physical volume that contains a mass
equal to the mass of the final halo at each redshift, assuming that
the sphere is at the mean density. To account for local emission, this
additional emissivity is added to the mean emissivity computed from
the global distribution of galaxies. We find that this local emission
is the dominant contribution to the net background at early times,
particularly prior to reionization when the global background has been
unable to build up as a result of the high optical depth in the
IGM. At later times, e.g. for $z<2$, the local contribution becomes
entirely negligible. Accounting for the local emission in this way is
an approximation that will be valid when the integrated background is
dominated by recently emitted light. Prior to reionization, this is
certainly the case, as emitted light is rapidly absorbed by neutral
gas. Post reionization, the local contribution to the emissivity
becomes small compared to the mean global emissivity and so can be
neglected.

Fig. ~\ref{fig:IGM_B06} shows the evolution of the proto-Local Group
region in the default feedback \citep{bower06} model, adopting an
escape fraction of ionizing photons of 100\%. The left-hand panel
illustrates the redshift evolution of the mean gas temperature in the
models with and without the local UV flux. At high redshift, the gas
cools adiabatically due to the expansion of the Universe. As the first
galaxies form, they begin to photoheat the IGM resulting in a rapid
rise in temperature at $z\simeq 20$ when the gas temperature soon
exceeds that of the CMB by over 2 orders of magnitude. In the model
with both global and local photoionization, the proto-Local Group
region is reionized at $z
\sim 14$. At that point, the gas reaches its maximum temperature of
just under $3\times 10^4$K. In the model without the contribution from
local sources, reionization occurs significantly later, at $z \sim 6$
(as in the previous calculation of \citealt{benson02a}) and the
maximum temperature that the gas attains is only $2\times
10^4$K. After reionization, the temperature of the proto-Local Group
gas declines slowly, at a rate controlled by the balance between
cooling and continued photoheating.

The right-hand panel of Fig.~\ref{fig:IGM_B06} shows the
characteristic accretion temperature of the gas, $T_{\rm accr}$, as a
function of redshift \citep[see][]{okamoto08}. The accretion
temperature rises rapidly at the epoch of reionization as a strong
photoheating background builds up, reaching a value of $\sim 8 \times
10^4$K at $z\simeq 3$ in the model with local sources. In the model
without local sources, the maximum temperature is approximately half as
high, $4.5 \times 10^4$K, and this is only reached at $z\simeq 1.5$. At
later times, the accretion temperature begins to fall as the
photoheating background declines.

The left panel of Fig.~\ref{fig:IGM_fbksat} shows the evolution of the
mean temperature of the proto-Local Group region in the model with
saturated feedback and global reionization only, for various assumed
escape fractions of ionizing photons. As expected, for a lower escape
fraction global reionization occurs later, but the model results in a
more plausible reionization history for large values of the escape
fraction
\citep[see][for a discussion of recent observational data]{stark10}. The 
earliest redshift of reionization in this case is $z\simeq 7.8$, for
$f_{\rm esc}=1$, and the latest is $z\simeq 4.5$, for $f_{\rm esc}=0.1$. The
peak of $T_{\rm protoLG}$ is similar in the two feedback models (compare
the solid blue lines in the left panels of Figs.~\ref{fig:IGM_B06}
and~\ref{fig:IGM_fbksat}). For $f_{\rm esc}=1$, this temperature is
$\sim\! 2\times 10^4$K. At a fixed escape fraction, global
reionization occurs earlier in the model with saturated feedback than
in the model with the default feedback of \cite{bower06} ($z\simeq
7.8$ compared with $z\simeq 6$ for $f_{\rm esc}=1$). This is because a
weaker feedback efficiency allows dwarf galaxies to become brighter
and therefore emit a larger number of ionizing photons.

The right panel of Fig.~\ref{fig:IGM_fbksat} shows the effect of
adding local reionization in the model with saturated feedback. (We
illustrate this for $f_{\rm esc}=1$, but the other cases display a similar
behaviour). In this case, local emission shifts the redshift of
reionization from $z\simeq 7.8$ to $z\simeq 10$, while the peak of the
gas temperature remains roughly the same. Below, we discuss how
different feedback and reionization models affect the satellite
luminosity function. 

\section{RESULTS}
\label{sec:results}

\begin{figure}
\includegraphics[width=9.cm] {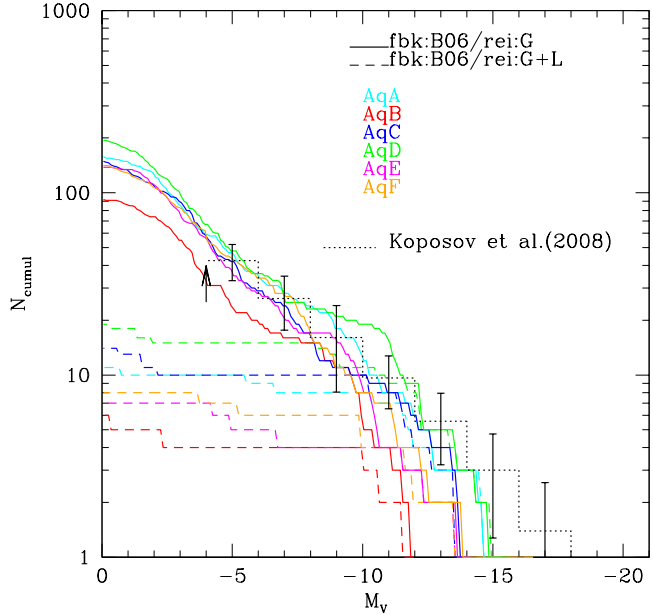}
\caption{\label{fig:lf_B06}{Luminosity functions for the models with
default feedback (full lines for global reionization: fbk:B06/rei:G;
dashed lines for global plus local reionization: fbk:B06/rei:G+L). The
different colours correspond to the six Aquarius halos. 
The estimate of the Milky Way luminosity function by
\citet{koposov08} is shown by the stepped dotted line down to the
limit of $M_{\rm V}=-5$, below which the volume corrections become
very uncertain (this limit is marked in the figure by a vertical
arrow). The errorbars are Poissonian for the satellites in the
classical ($M_{\rm V} \le -11$)} regime, while for the fainter systems they
include volume corrections appropriate for the SDSS DR5.}
\end{figure}

In this section, we investigate how different feedback processes
affect the broad properties of satellite galaxies. We compare our
model predictions not only to the observed satellite luminosity
function\footnote{Unless stated otherwise, we use the term `luminosity
function' to refer to the luminosity function of dwarf satellite
galaxies around galaxies like the Milky Way.}, but also to the
observed satellite metallicity-luminosity relation. As we shall see,
this relation serves to break a degeneracy in the models between the
effects of supernova feedback and reionization.

We will compare our model predictions for the luminosity function with
data for the Milky Way, based largely on the SDSS DR5 data.  For this,
we count all satellites in the model within a radius of 280~kpc, the
limit to which the tip of the red giant branch can be detected in the
SDSS. We compared this luminosity function to the estimate by 
\citet{koposov08} of the expected number of satellites in the Milky
Way within this radius, which is well described by a power-law,

\begin{equation}
dN/dM_{\rm V} = 10 \times 10^{0.1(M_{\rm V} + 5)}. 
\label{eq:Kopfit}
\end{equation}

\noindent The  number of satellites in   the Milky Way estimated
according to this calculation is shown in the luminosity function
plots below by a stepped dotted line. In the ``classical dwarfs'' regime
($M_{\rm V} \leq -11$), the error bars on the observational estimate
are assumed to be Poissonian, while in the interval $-11 \le M_{\rm
V} \le -5$, additional corrections for the SDSS DR5 volume are
included. We do not plot the observational estimates for magnitudes
fainter than $M_{\rm V}=-5$ where the volume corrections become very
uncertain. Note that in this cumulative plot the errors are not
independent. Also note that the above estimate requires an uncertain
assumption about the radial distribution of Milky Way satellites
\citep{tollerud08}.

\begin{figure*}
\includegraphics[width=8.cm]{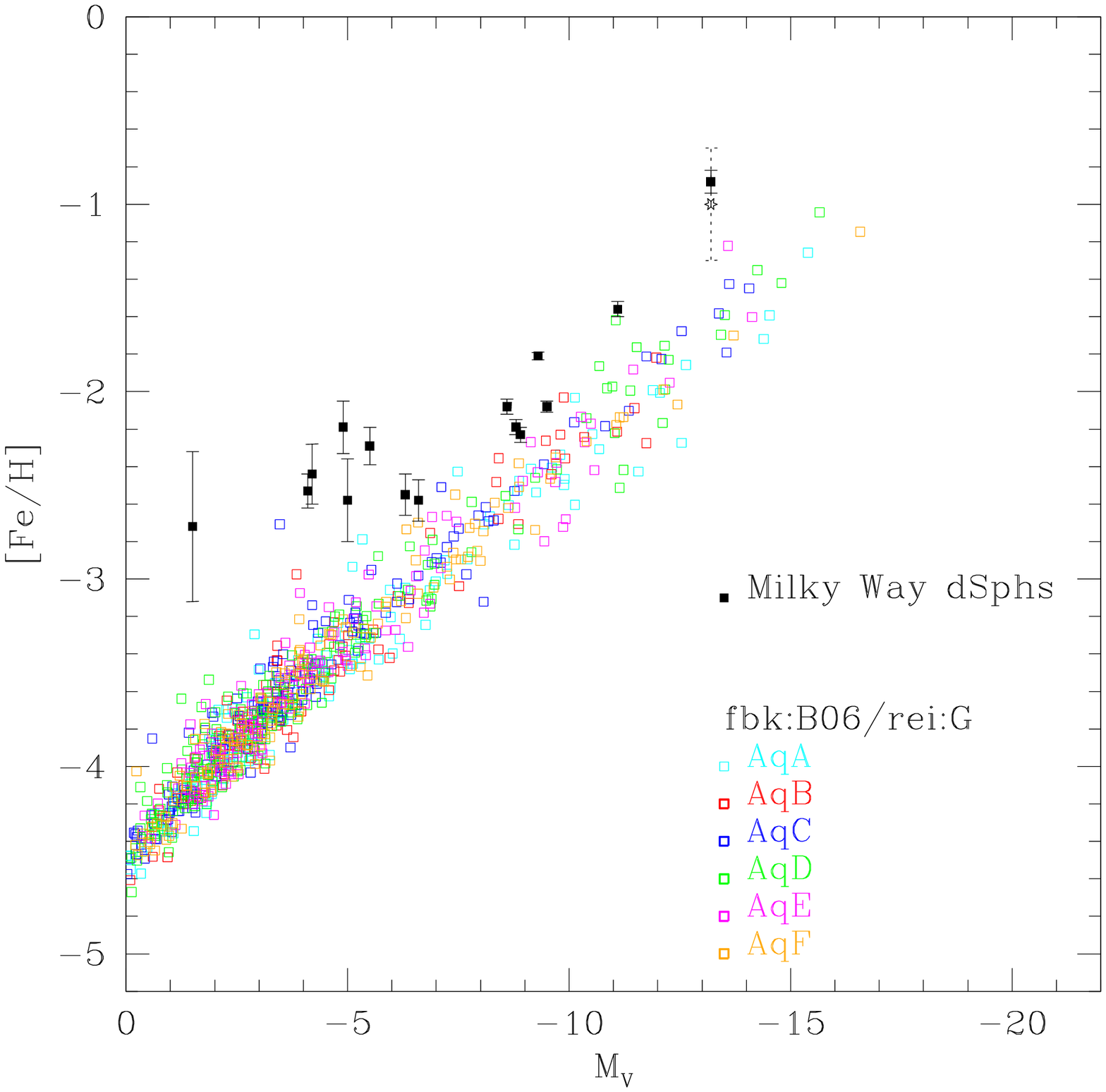}
\includegraphics[width=8.cm]{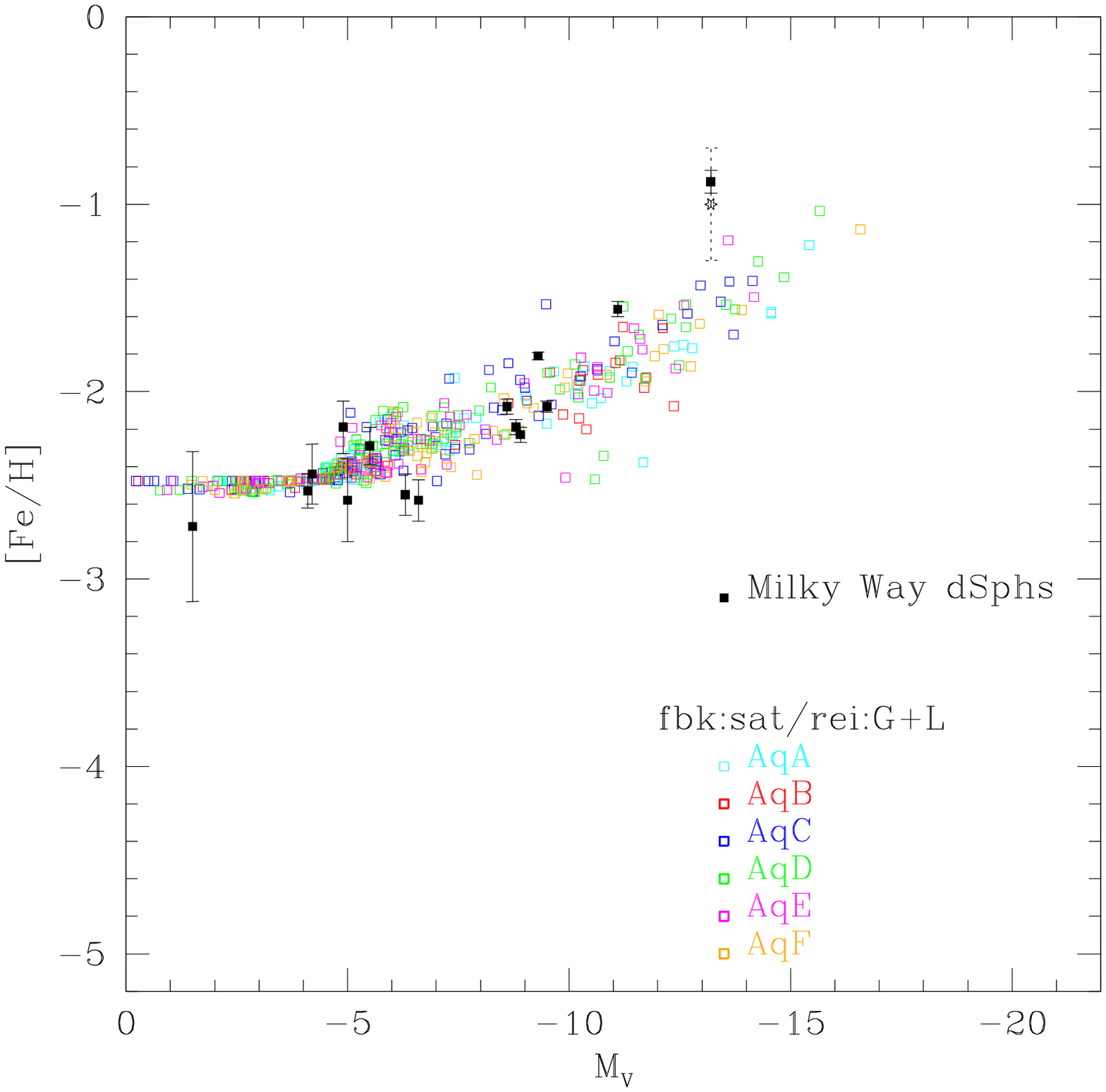}
\caption{\label{fig:feh}{The $Z - M_{\rm V}$ relation for the
fbk:B06/rei:G models ({\it left}) and for the fbk:sat/rei:G+L models 
({\it right}), where $Z$ is the average [Fe/H].  Both models assume 
$f_{\rm esc}=1$. The different colours correspond to satellites in the 
six Aquarius halos. The black squares show the observational data for 
the Milky Way dwarf spheroidals \citep{norris10}. This study includes 
data for both classical dwarfs 
(\citealt{mateo98,helmi06}) and ultra-faint dwarfs 
(\citealt{kirby08,martin08,norris10}), but only for those systems 
with chemically unbiased star samples. The [Fe/H] error bars have been 
re-calculated by \citet{norris10} using the precepts of \citealt{dacosta77} 
(\S~III). The star symbol shows the peak of the metallicity distribution 
function in Fornax measured by \citet{battaglia06b} and \citet{kirby10} using a
larger sample of stars but with lower resolution spectra than 
\citet{norris10}. The lower resolution samples cover a larger spatial 
extent in Fornax and therefore may give a more accurate representation of 
the average [Fe/H] by taking into account the contribution of 
metal-poor populations preferentially located at the outskirts 
\citep{battaglia06b,letarte10}.}}
\end{figure*}

For properties of the satellite population for which the
radial incompleteness of the observational data is important
(e.g. the radial distribution of satellites in Section~\ref{sec:rad}), we
model the incompleteness explicitly within the SDSS DR5
footprint. Firstly, we apply the magnitude dependent 
threshold of detectability, $R_{max}(M_{\rm V})$, derived by
\citet{koposov08}. Then we randomly select $20\%$ of the satellites to 
account for the SDSS DR5 partial coverage of the sky. The model ``classical''
satellites ($M_{\rm V} \le -11$) are all included, under the
assumption that they are not affected by incompleteness.

\subsection{The default feedback model (fbk:B06)}
\label{sec:sf}

The combined effects of supernova feedback and reionization in the
default feedback model, applied to the six Aquarius halos, are shown in
Fig.~\ref{fig:lf_B06}. Models with global reionization only are
labelled fbk:B06/rei:G (solid lines) and models which, in addition, 
include local reionization are labelled fbk:B06/rei:G+L (dashed lines).

Within the scatter, the fbk:B06/rei:G models give a reasonable match
to the luminosity function. However, none of the halos host galaxies
as bright as the the SMC and the LMC. A similar result was obtained by
\citet{benson02b}, who found that only about 1 in 20 of their 
Milky Way-type galaxies had satellites as bright as the SMC and the
LMC. Remarkably, this result now appears consistent with recent
measurements of the prevalence of bright satellites in external
galaxies similar to the Milky Way which indicate that only 11\% of
such hosts have one and 3.5\% two satellites as bright as the
Magellanic Clouds \citep{liu11}. Note that there is a large variation
in the number of model satellites from one halo to another: the
predicted satellite abundance varies by a factor of about~2,
reflecting the different formation histories of the halos. This
relatively large scatter highlights the danger of arriving at
far-reaching conclusions regarding cosmology based on the single
example of the Milky Way. Indeed, \cite{guo11b} have recently shown, 
using SDSS data, that, in the mean, isolated primaries of comparable
luminosity to the Milky Way contain about a factor of two fewer
satellites brighter than $M_V=-14$ than the Milky Way itself.

In the models, there is a degeneracy between the effects of supernovae
feedback and reionization: both suppress galaxy formation in small
halos. The metallicity, $Z$, of the stars and gas in a galaxy that is
already assembled is not affected by photoheating. It can, however, be
strongly affected by supernova feedback which reduces the effective
yield as a consequence of outflows.  Thus, the $Z - M_{\rm V}$
relation has the potential to break this degeneracy.

\begin{figure*}
\begin{tabular}{c |c }
\includegraphics[width=9.cm]{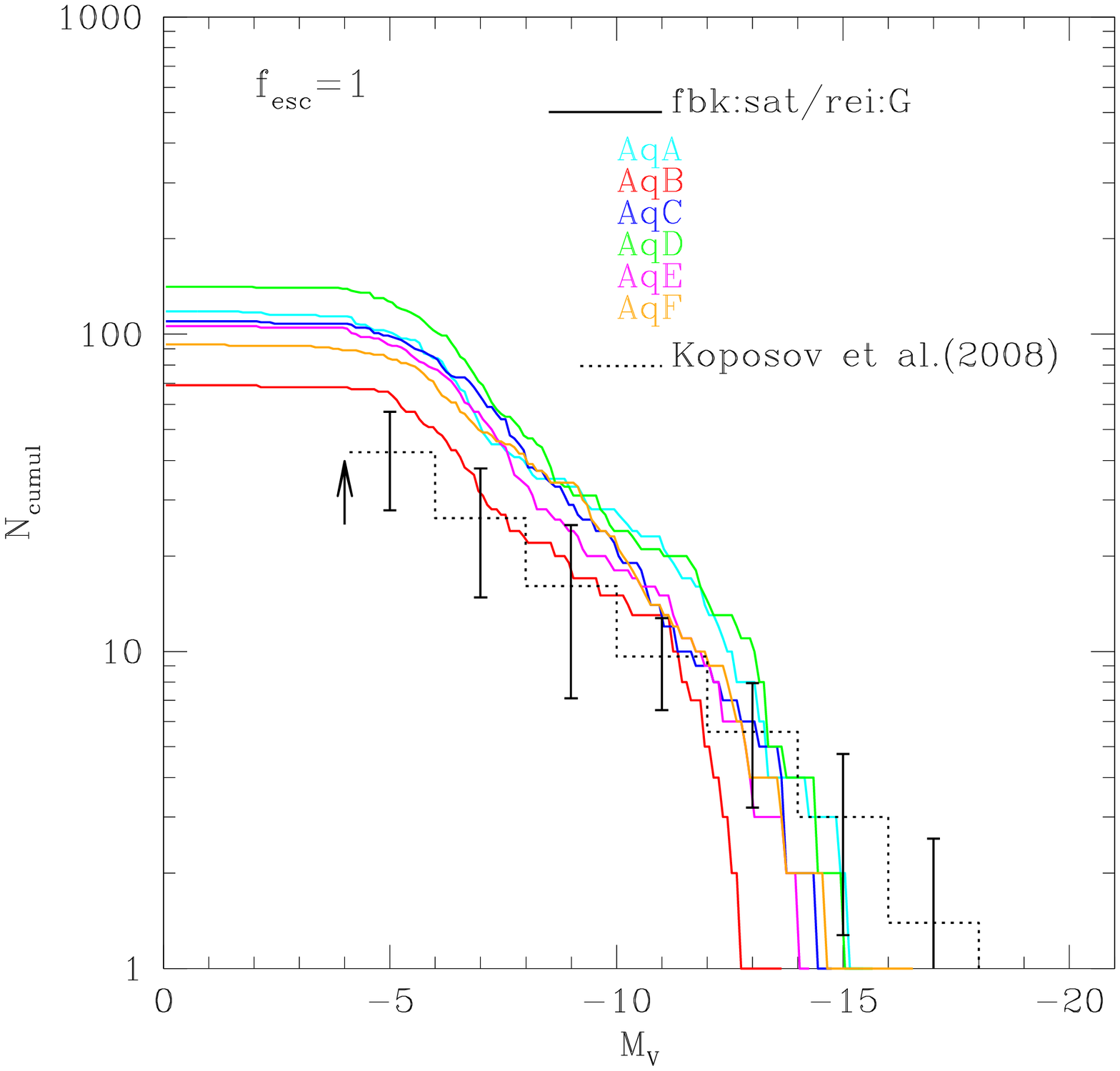} &
\includegraphics[width=9.cm]{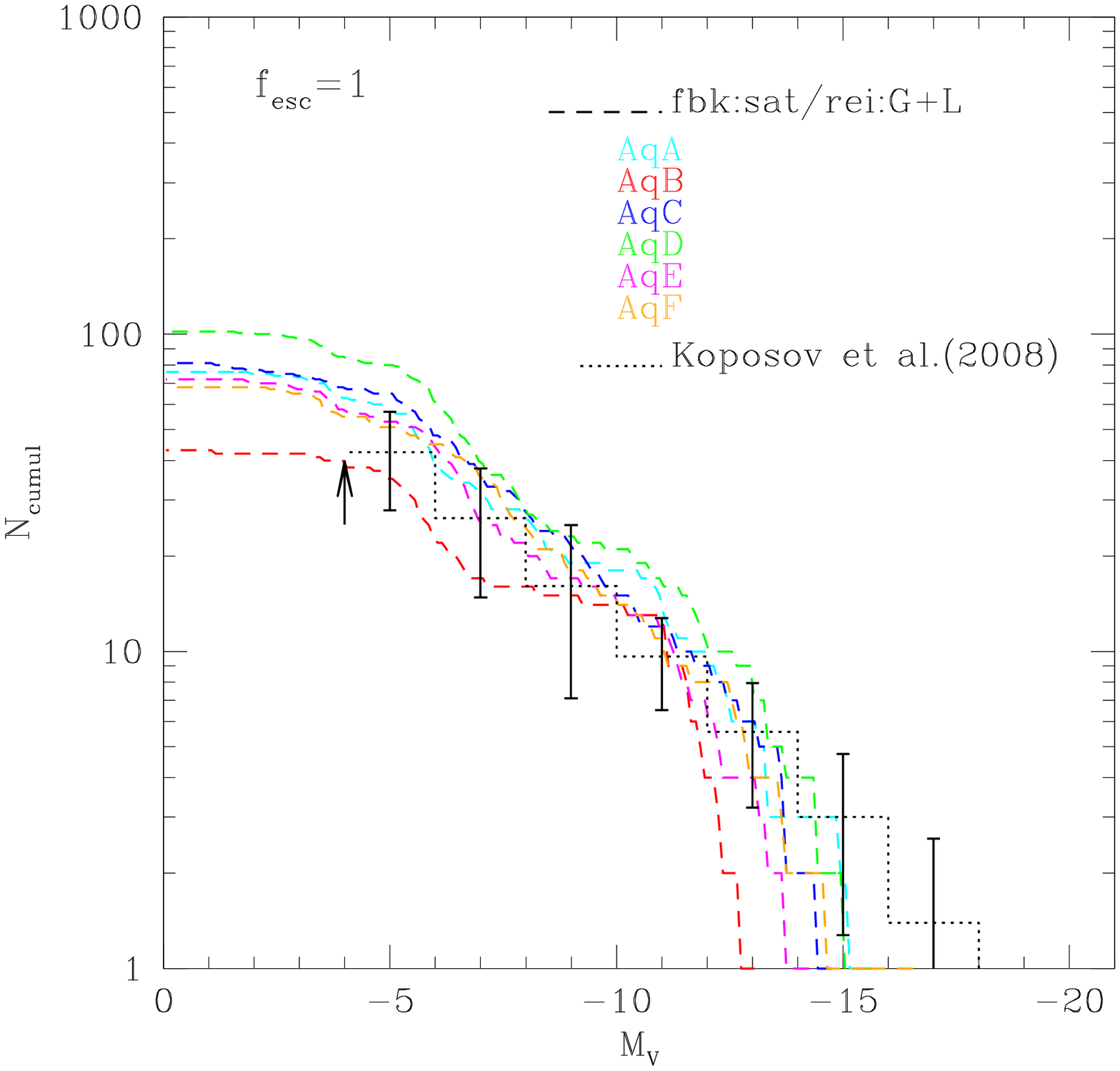} \\
\end{tabular}
\caption{\label{fig:lf_fbksat}{Satellite luminosity functions in the
fbk:sat/rei:G and fbk:sat/rei:G+L models for the six Aquarius
halos. The photon escape fraction is taken to be $f_{\rm esc}=1$. The
solid lines represent the models with global reionization only and the
dashed lines the models with both local and global reionization;
different lines correspond to the six Aquarius halos. The
observational estimate of \citet{koposov08} is shown as a stepped
dotted line, as in Fig.~\ref{fig:lf_B06}. }}
\end{figure*}
 
The $Z - M_{\rm V}$ relation in the fbk:B06/rei:G model is shown in
the left panel of Fig.~\ref{fig:feh}, for all six Aquarius halos, and
compared with observations of Milky Way dwarfs with reliable
metallicity measurements \citep{norris10}. The 
default feedback model undershoots the metallicities of dwarf
galaxies, particularly for the less luminous systems. (Note that the
model already includes the large yield, $p=0.04$, favoured by
\cite{font08}). The scatter in the $Z - M_{\rm V}$ relation for all
six Aquarius systems is relatively small, suggesting that the merger history
plays only a secondary role in determining the slope or zero-point of
the relation. We conclude that the supernova feedback in this model is
too efficient at expelling metals from dwarf galaxies. We now analyze
the saturated feedback model.

\subsection{The saturated feedback model (fbk:sat)}
\label{sec:fbksat}

The right panel of Fig.~\ref{fig:feh} shows the $Z - M_{\rm V}$
relation in the model fbk:sat/rei:G+L. With this feedback
prescription, the match to the observed $Z - M_{\rm V}$ relation is
greatly improved, especially for the fainter satellites. In
particular, the turnover in satellite metallicities is well
reproduced. As discussed in Section~\ref{sec:basic}, a floor in the average
metallicity of satellites can be created by a very inefficient
supernova feedback\footnote{We could not reproduce this flattening of
$Z$ at faint $M_{\rm V}$ with a typical power-law supernova feedback,
regardless of the change in slope or zero-point.}. 
There are two main regimes in the evolution of dwarf galaxies 
with inefficient feedback. At early times, there is a significant gas 
inflow that sustains star formation and an increase metallicity up to 
$\sim ~10^{-2.5}Z_{\odot}$ (a single generation of supernovae may be sufficient
to increase the gas metallicity by $\sim 0.5$~dex). This regime is short for 
the smaller dwarf galaxies, as their gas inflow drops quickly. From then on, 
the rate of change in metals becomes directly proportional with the 
outflow which, assuming a constant feedback efficiency, leads to a 
floor in metallicity.
 
At the brightest end (i.e., Fornax) our model predicts average
metallicities that are below the data point of
\citet{norris10}. However, this measurement is likely to be an
overestimate of the actual value because the sample used
preferentially contains stars close to the centre of the galaxy (which
have higher  resolution spectra), and here the stars are more
metal-rich \citep{letarte10}. Our data are in better agreement with
the average $Z$ values inferred from lower resolution spectra that
cover a larger extent of this dwarf galaxy
\citep{battaglia06b,kirby10}.

Fig.~\ref{fig:lf_fbksat} shows the luminosity functions of the six
Aquarius halos in the fbk:sat/rei:G and fbk:sat/rei:G+L models, for an
assumed photon escape fraction of $f_{\rm esc}=1$. This high value is
required to match the observed luminosity function. As mentioned
earlier, this value produces a plausible reionization history, but
seems higher than indicated by some observational data. Although the
escape fraction remains uncertain, such high values (which arise in
\galform\, because too few ionizing photons are produced at high
redshifts) is probably a shortcoming of the model and requires further
investigation. However, for our purposes here, the actual value of the
escape fraction is not in itself important. What matters is that there
should be enough ionizing radiation to suppress the formation of small
galaxies.

Fig.~\ref{fig:lf_fbksat} indicates that, unlike in the fbk:B06 case,
global reionization alone does not suppress galaxy formation enough in
the fbk:sat model to account for the satellite luminosity function. In
this case, reionization by local sources is also required.  There are
important differences in our two feedback models. In particular, the
saturated feedback model produces more satellites of all luminosities,
even with $f_{\rm esc}=1$, than the default \cite{bower06} feedback model
(compare Figs.~\ref{fig:lf_B06} and~\ref{fig:lf_fbksat}.) This is
largely because the lower feedback efficiency in the saturated model
allows galaxies to grow brighter.

\begin{figure*}
\includegraphics[width=\columnwidth]{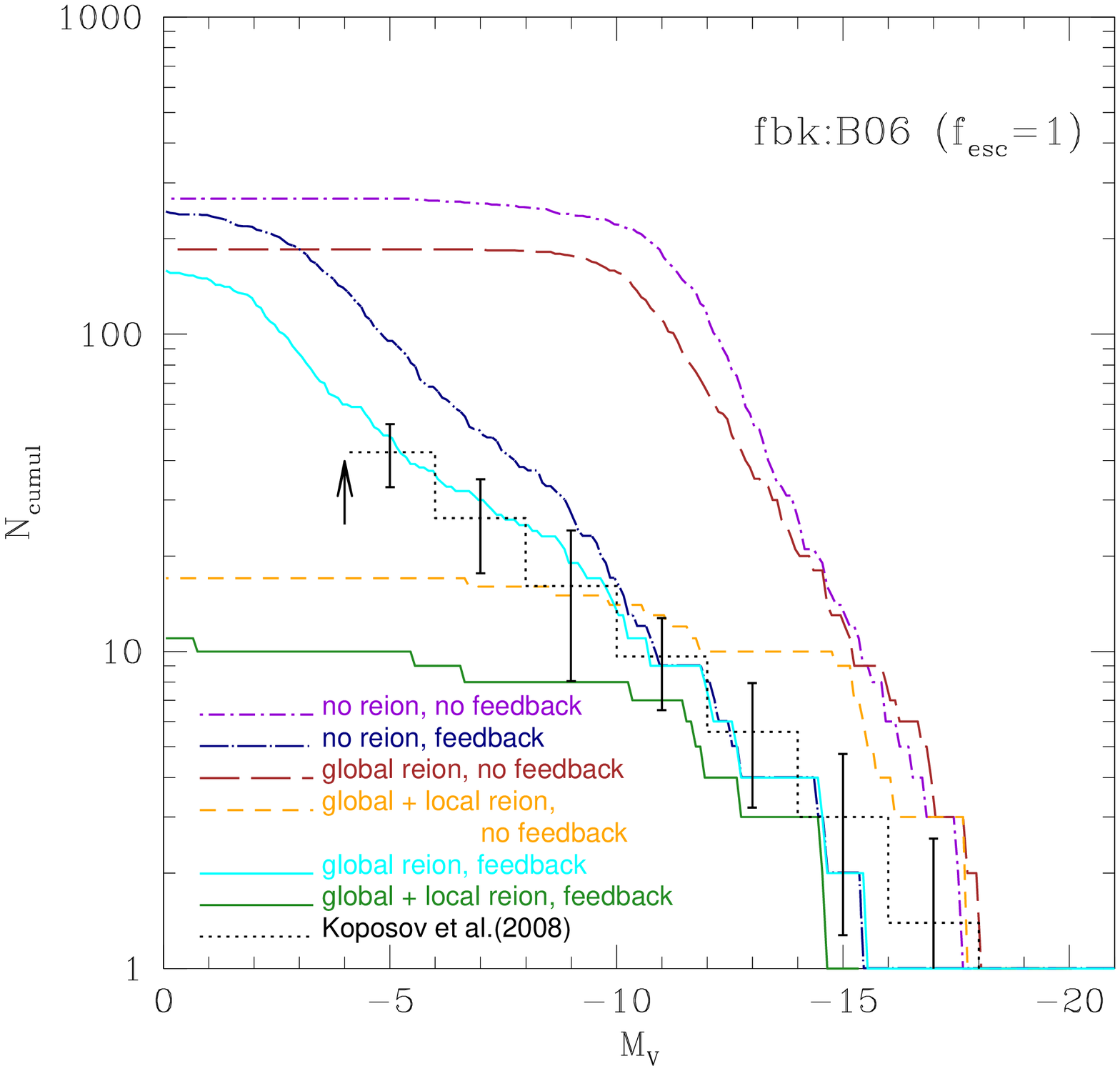}
\includegraphics[width=\columnwidth]{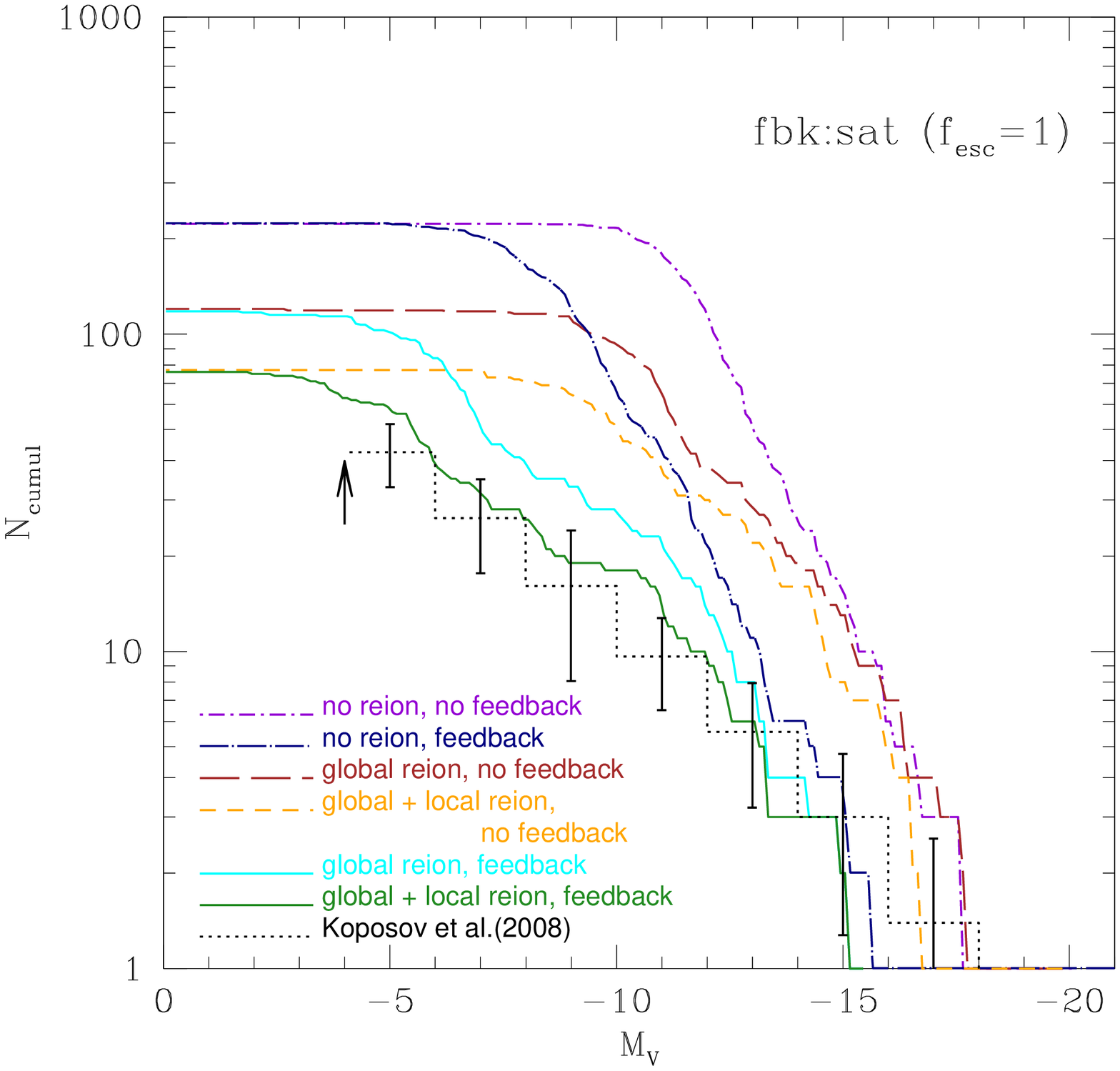}
\caption{\label{fig:lf_reion_fdbk}{The effect of different physical
processes on the luminosity function of satellites in the Aq-A
halo, in the models fbk:B06 ({\it left}) and fbk:sat ({\it right}). 
Both models shown here have $f_{\rm esc}=1$. The purple short-long dashed 
line shows a model without any reionization or supernova feedback. 
The dark blue dot-long dashed line shows a model with supernova 
feedback only, no reionization.  The two pure dashed lines show models 
without supernova feedback but with reionization, dark red long dashes for 
global reionization only and orange short dashes for global plus local 
reionization. The cyan full line shows the model with global reionization 
and feedback. The model with global plus local reionization and supernova 
feedback is shown with full dark green lines.  
The observational data and their errorbars are as in Fig. \ref{fig:lf_B06}.}}
\end{figure*}

In summary, we have found that the model fbk:sat/rei:G+L matches both
the global luminosity function (of all galaxies; see
Fig.~\ref{fig:global_KLF}), the local luminosity function of Milky Way
satellites (see Fig.~\ref{fig:lf_fbksat}) and the $Z - M_{\rm V}$
relation (see Fig.~\ref{fig:feh}). This model also produces central
galaxy stellar masses roughly consistent with the measured stellar
mass of the Milky Way (see Table~\ref{table:Aq}). In
Section~\ref{sec:tests} we will adopt this as the fiducial model and show
that it also matches other properties of the Milky Way dwarf
satellites.

\subsection{The effects of supernova feedback and reionization}

Before exploring our fiducial model further, it is instructive to
compare the separate effects of the two sources of feedback, supernova
energy and reionization, on the satellite luminosity function in the
two cases, fbk:B06 and fbk:sat. These are illustrated in the two
panels of Fig.~\ref{fig:lf_reion_fdbk}.

In the absence of feedback of any kind, the purple dot-dashed lines
show that in both cases hundreds of satellite galaxies form with
$V$-magnitudes brighter than $M_{\rm V}= -10$. The lack of feedback
enables galaxies to retain their baryons and continue to grow in size
and luminosity. Below $M_{\rm V}\approx -10$, the cumulative
luminosity function levels off due to the inability of gas in halos
with virial temperatures below $10^4$K to cool efficiently. The effect
of reionization by a global UV background (dark red long dashed lines)
is not enough, on its own, to lower the number of satellites
sufficiently so as to match the observed luminosity function. The
effect is minimal in the fbk:B06 model, where it suppresses the
formation of satellites only by about 50\%. In the fbk:sat model it
reduces the number of luminous satellites by a factor of 2, but still
overpredicts the bright end. Our model of reionization is based on the
hydrodynamic simulations of \citet{okamoto08} in which the effects are
relatively mild, with gas accretion being suppressed only in halos
with circular velocity lower than $\sim 25\ {\rm km\ s}^{-1}$. Most
previous studies of satellite galaxies have assumed the more
aggressive reionization model of \citet{gnedin00} in which suppression
occurs in halos with circular velocity up to $\sim 50\ {\rm km\
s}^{-1}$. However, even in this case, the effects of global
reionization are relatively mild \citep[see e.g.][]{benson02b}.

The inclusion of local sources of reionization (orange short dashed
lines in Fig.~\ref{fig:lf_reion_fdbk}) has a dramatic effect in
fbk:B06, but a relatively weaker effect in the fbk:sat model. This is
to be expected since in the fbk:sat model a larger fraction of
ionizing emission arises from lower mass progenitors. In the overdense
region corresponding to the proto-Local Group these low mass
progenitors are less overabundant (relative to their mean abundance
averaged over all space) than are more massive progenitors
\citep{mowhite96}. Thus, the enhancement in the total ionizing
emissivity is similarly reduced when accounting for their local
contribution. In both cases reionization is pushed back locally to an
earlier redshift and galaxy formation is inhibited in halos with very
low velocity dispersion. The bright end, $M_{\rm V}<-15$, remains
unaffected.

Perhaps the most illuminating is the effect changing the supernova
feedback. In the fbk:B06 model this process affects galaxies in halos
of all circular velocities. The efficiency is high and supernova
feedback dominates global reionization across the entire range of
satellite galaxies. In the fbk:sat model, this process becomes
scale-dependent. Interestingly, the scale on which supernova feedback
saturates, $M_{\rm V} \approx -10$, it is also where it becomes less
efficient than global reionization in suppressing the formation of
satellites. Feedback alone (shown by dark blue dot-dashed lines)
cannot explain the luminosity function in either model: the abundance
of galaxies overshoots the data by a factor of $\sim 2$ in the fbk:B06
case and by a factor of $\sim 10$ in the fbk:sat case. In both cases,
supernova feedback is the dominant source of suppression of massive,
bright satellites (with $M_{\rm V} < -10$). As found in earlier
studies, a combination of feedback and photoheating (global only in
fbk:B06 and global + local in fbk:sat) is required to account for the
abundance of low-mass satellites.

\section{Tests of the fiducial model}
\label{sec:tests}

We have selected the parameters of the fiducial model 
(fbk:sat/rei:G+L) on the basis of the luminosity function and 
$Z - M_{\rm V}$ of Milky Way satellites. In this section, we test the model
against measurements of the radial distribution of satellites and the relation 
between the central mass density and luminosity. It is important to note that
the model has not been adjusted during this comparison.

\subsection{The radial distribution of satellites} 
\label{sec:rad}

\begin{figure}
\includegraphics[width=9.cm]{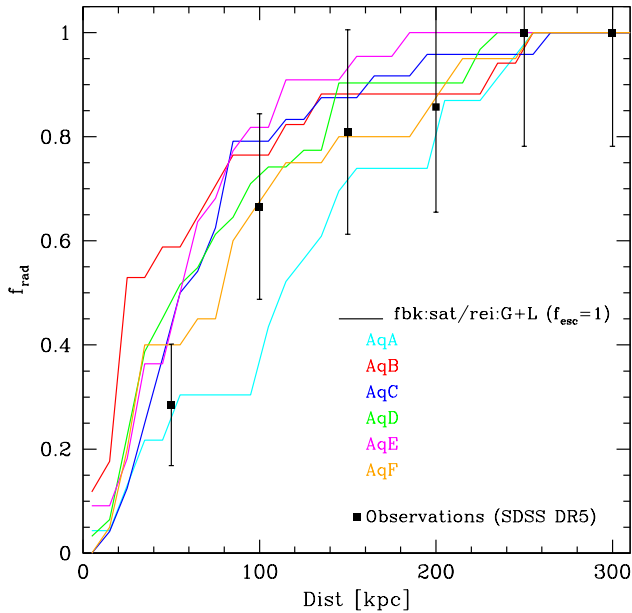}
\caption{\label{fig:rad_df}{Radial distribution of satellites in the 
fiducial fbk:sat/rei:G+L  model. The observations include both the
 11 classical satellites and those detected in SDSS DR5 (compiled from 
\citet{mateo98} and \citet{koposov09}), for which the SDSS DR5 threshold 
selection and sky coverage has been assumed (see text for
details). The error bars illustrate the uncertainty due to Poisson
statistics. Note that in this cumulative plot the errors are not
independent.}}
\end{figure}

Fig.~\ref{fig:rad_df} shows the radial distribution of the satellites
that survive to the present day in the fiducial model for the six
Aquarius halos. As outlined in Section~\ref{sec:results}, for magnitudes
fainter than the classical dwarf regime ($M_{\rm V} \ge -11$), we
select only those satellites that could have been observed in the SDSS
DR5 survey. The detection of brighter satellites is assumed to be
complete and so these are all included. Satellite distances and their
uncertainties (typically 10-20\%) are taken from \citet{mateo98} in
the case of the classical dwarfs and from \citet{koposov09} in the
case of the ultra-faint dwarfs detected in the SDSS DR5.

Overall, the new model displays a radial distribution of surviving
satellites similar to the observations (which have large
uncertainties). The predicted distributions appear slightly more
concentrated than the data for 4 of the 6 halos, less concentrated
than the data for one and very close to the data for the remaining
halo.

\subsection{The $M_{300} - L$ relation}
\label{sec:m300lum}

Dwarf galaxies are inefficient retainers of baryons and, as a result,
are strongly dark matter dominated. This makes them ideal probes of
the dark matter. In particular, their very central regions could
contain information about its identity
\cite[e.g.][]{nfw96}. Rough estimates suggest 
a mass of about $10^{7}$~M$_{\odot}$ within the visible parts of the
Milky Way satellite \citep{mateo98,gilmore07}. More robust analyses
give total masses of about $3\times 10^{9}$ M$_{\odot}$
\citep{wolf10,walker10}, and, surprisingly, a common mass contained within the
central 300~pc, $M_{300} \sim 10^{7}$~M$_{\odot}$, independently of
luminosity over four orders of magnitude \citep{strigari08}.

\begin{figure*}
\includegraphics[width=18.cm]{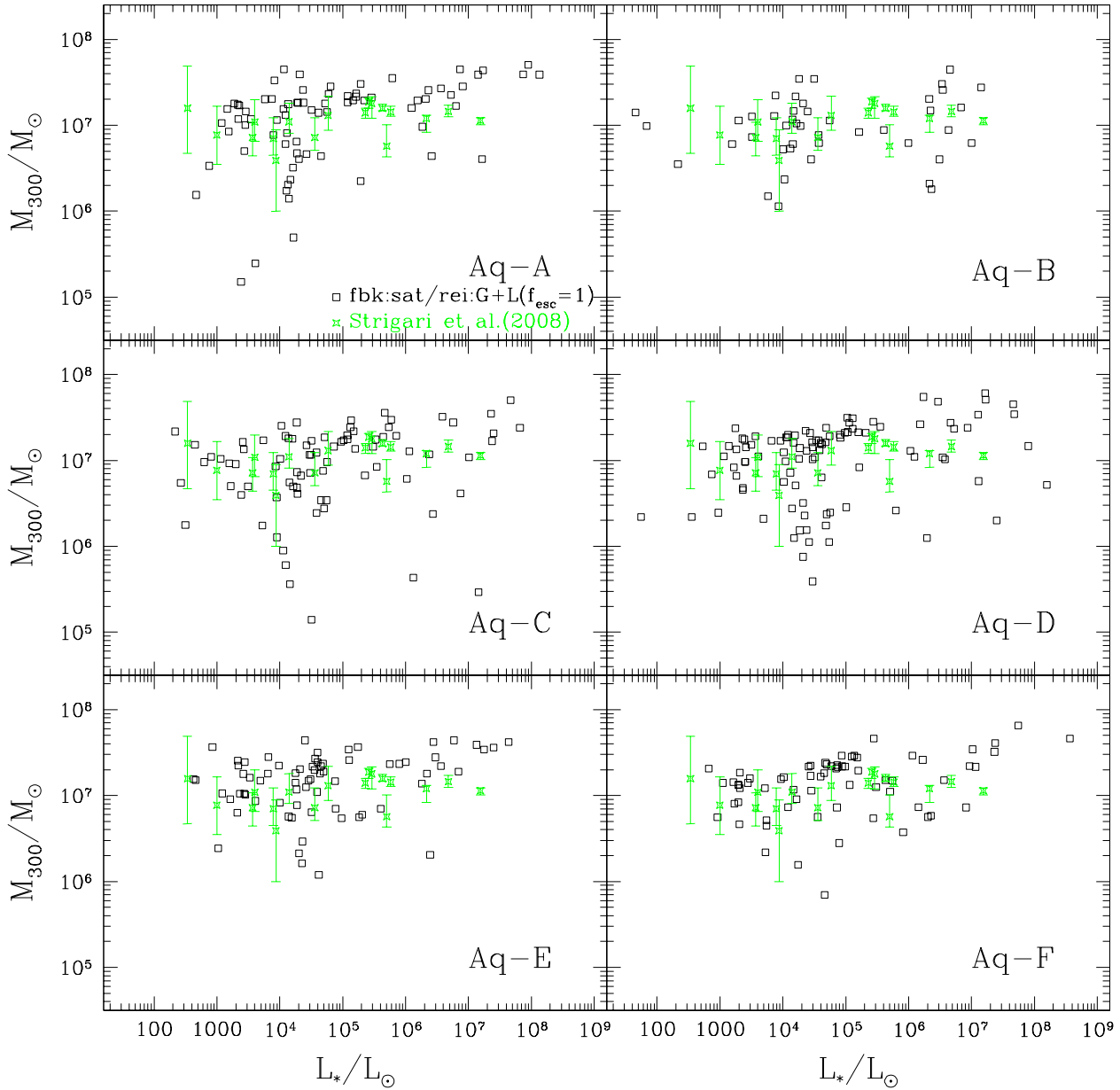}
\caption{\label{fig:m300_lum}{The $M_{300} - L$ relation for
satellites in the six Aquarius halos. Results for the fiducial model are 
shown as empty squares. Observational estimates of $M_{300}$ (green stars) 
are from \citet{strigari08}.}} 
\end{figure*}

\begin{figure*}
\includegraphics[width=18.cm]{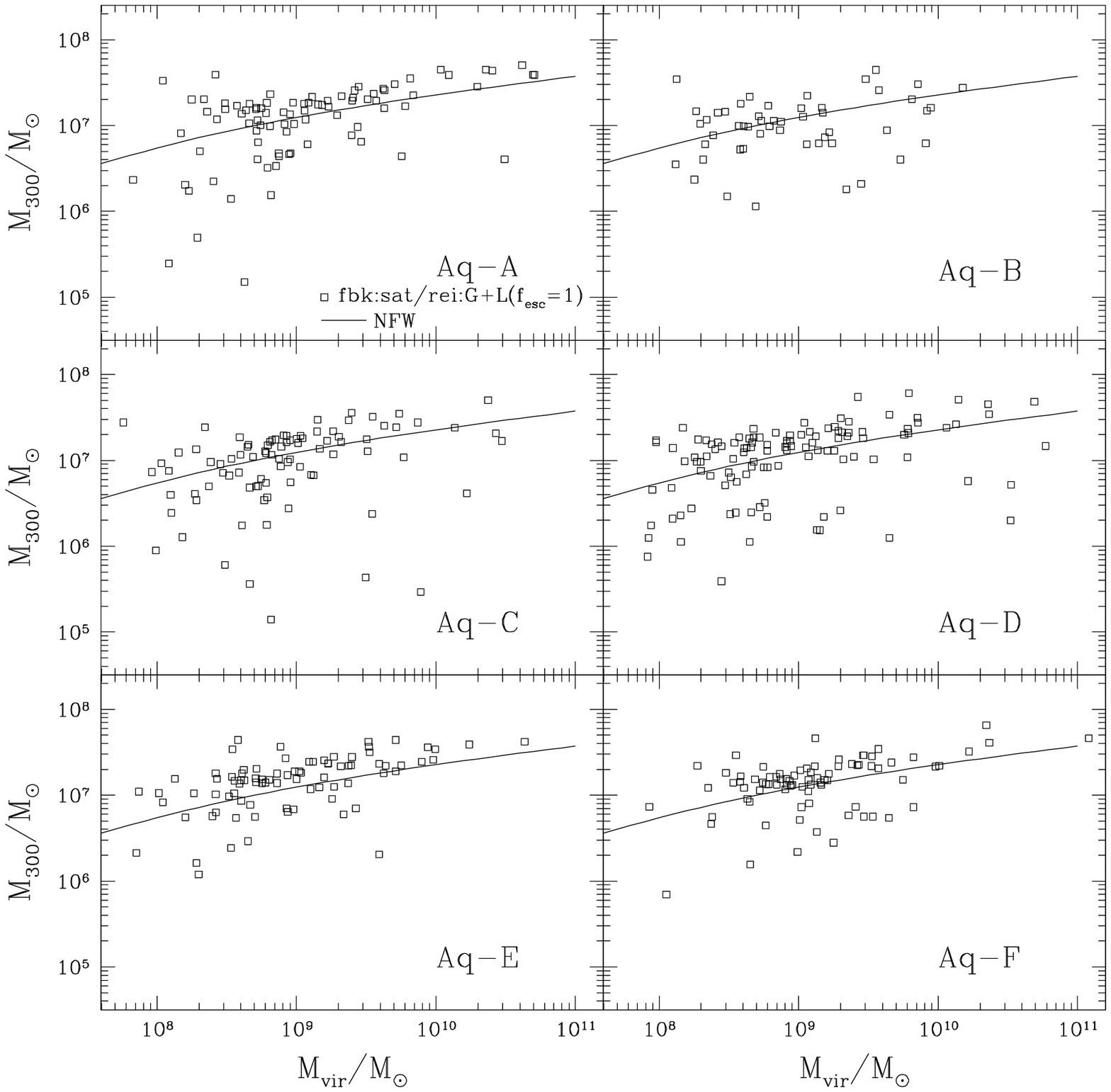}
\caption{\label{fig:m300_mvir}{The $M_{300} - M_{\rm vir}$ relation for 
satellites in the six Aquarius halos. Results for the fiducial model are 
shown as empty squares. The solid line in each panel shows the $M_{300} - M_{\rm
vir}$ relation determined from the \citet{nfw97} dark matter profile,
using the mean mass $-$ concentration relation found by
\citet{neto07}.}}
\end{figure*}

The high numerical resolution of the Aquarius halos allows us to
calculate $M_{300}$ directly from the simulations and, in conjunction
with our modelling of galaxy formation in subhalos, to determine how
it varies with satellite luminosity. 
Fig.~\ref{fig:m300_lum} shows the $M_{300} - L$ relation for the
satellite galaxies in the six simulated Aquarius halos, in the case of 
the fiducial model. We compute $M_{300}$ directly from the simulations, 
by summing the masses of all particles within 300~pc of the centre of 
each subhalo. 

Since 300~pc is close to the resolution limit of the Aquarius
simulations \citep{springel08b}, we have checked the numerical
convergence of $M_{300}$ using halo Aq-A, for which a higher
resolution simulation, at level~1, is available. We created catalogues
of subhalos matched in the level~1 and level~2 simulations by pairing
up halos according to position, velocity and mass.  We find an
extremely good correlation between $M_{300}$ for the most massive
subhalos at the two levels of resolution, with more scatter present at
lower masses, as expected. However, we find that the level~2 subhalos
systematically underestimate $M_{300}$ relative to level~1 by about
20\%, even in the most massive systems. This difference is
sufficiently small that it will not affect our conclusions, but should
be kept in mind when comparing model predictions with data. In the 
following, we present only the level~2 simulations.

Fig.~\ref{fig:m300_lum} shows that there is broad consistency in
$M_{300}$ between the Aquarius simulations and the data of
\citet{strigari08}. However, the masses in the simulations have larger
scatter and, in some cases, slightly larger values than the
observations. In all cases the simulations show a weak trend with
luminosity that is not apparent in the data. The gasdynamic
simulations of galaxy formation in Aquarius halo~D by \cite{okamoto09}
show a similar trend between $M_{600}$ and luminosity.

In some of our models, a population of very faint 
satellites forms which have values of $M_{300}$ lower than the 
typical $\sim 10^{7}$ M$_{\odot}$ estimated for the real satellites. It may 
be that systems like this exist in nature but have not been detected so far
because of their intrinsically faint luminosities and low surface
brightnesses. At the bright end the simulations show a large scatter in
 $M_{300}$ above and below $10^7 M_\odot$. Better observational statistics 
may help understand the reason for this discrepancy. 

The physical origin of the $M_{300}$ -- $L$ relation has recently been
discussed by \cite{stringer10}. We explore this question here by examining
the dependence of $M_{300}$ on satellite virial mass.

\subsection{The $M_{300}-M_{\rm vir}$ relation}
\label{sec:m300mvir}

In Fig.~\ref{fig:m300_mvir} we plot the $M_{300} - M_{\rm vir}$
relation for satellites in the six Aquarius halos. Here, $M_{\rm vir}$ 
is the virial mass before infall into the main halo. The solid line 
in each panel shows the $M_{300} - M_{\rm vir}$ relation derived for 
dark matter halos with NFW density profiles (\citealt{nfw96,nfw97}), 
assuming the best-fit average mass~--~concentration relation found by
\citet{neto07}. The $M_{300} - M_{\rm vir}$ relation has a similar
shape to the $M_{300} - L$ relation, that is, $M_{300}$ depends only
weakly on halo virial mass. A power-law fit to the trend yields
$M_{300} \propto M_{\rm vir}^{1/3}$.  This result can be understood as
follows.  For a halo of mass $M_{\rm vir}=10^9 M_\odot$ in the
$\Lambda$CDM cosmology, the ratio of 300~pc to $r_{\rm vir}$ is
$\approx 0.015$ and scales as

\begin{equation}
\frac{r_{300}}{r_{\rm vir}} \approx 0.015 \biggl(\frac{M_{\rm
vir}}{10^9 M_\odot}\biggr)^{-1/3}. 
\end{equation} 

At small radii (in the limit of $r \ll r_{\rm s}$, where $r_{\rm s}$
is the NFW scale radius; \citealt{nfw97}) in a given halo, the NFW
density profile asymptotes to $\rho \propto r^{-1}$, implying a mass
profile $M(<r) \propto r^2$.  Since the radius $r_{300}$ is much
smaller than $r_{\rm s}$ for the halo masses under consideration, the
ratio $M_{300}/M_{\rm vir}$ scales as $(r_{300}/r_{\rm vir})^2$.
Substituting in the scaling for $r_{300}/r_{\rm vir}$ in equation (4)
yields the scaling $M_{300}
\propto M_{\rm vir}^{1/3}$ found in the Aquarius simulations.

The weak trend in the $M_{300}$--$M_{\rm vir}$ relation is the
underlying cause of the weak trend in the $M_{300}$--$L$
relation and appears to be a robust prediction of cold dark matter
theory unless baryonic processes in the forming dwarf galaxy are able
to modify the inner regions of the dark matter density profile
significantly.

\subsection{A parameterization of the $M_{300} - L$ relation}
\label{sec:param_m300lum}

Summing up the results above, we can now understand why $M_{300}$
appears to be independent of $L$, even if in reality it has some weak
dependence. According to our results, $L$ varies as ${M_{\rm
vir}}^{n}$, where $n \sim 2.5$, and $M_{300}$ varies as $M_{\rm
vir}^{1/3}$.  This yields the very weak dependence, $M_{300} \sim
L^{\beta}$, where $\beta \sim 0.1$, consistent with current
observations. This trend is a robust prediction of our models which
may, in principle, be tested by further photometric and spectroscopic
observations of the faint satellites of the Local Group.

Other theoretical studies find it similarly difficult to reproduce the
completely flat $M_{300}-L$ relation seen in the data
\citep{koposov09,maccio09,li09,munoz09,okamoto09,busha10,stringer10}.  In particular,
\citet{munoz09} find a similar fit to ours ($M_{300} \sim L^{0.22}$) using
the $M_{300}$ values derived from their simulation.

\subsection{Comparison with the Li et al. (2010) model}
\label{sec:li}

\begin{figure}
\includegraphics[width=9.cm] {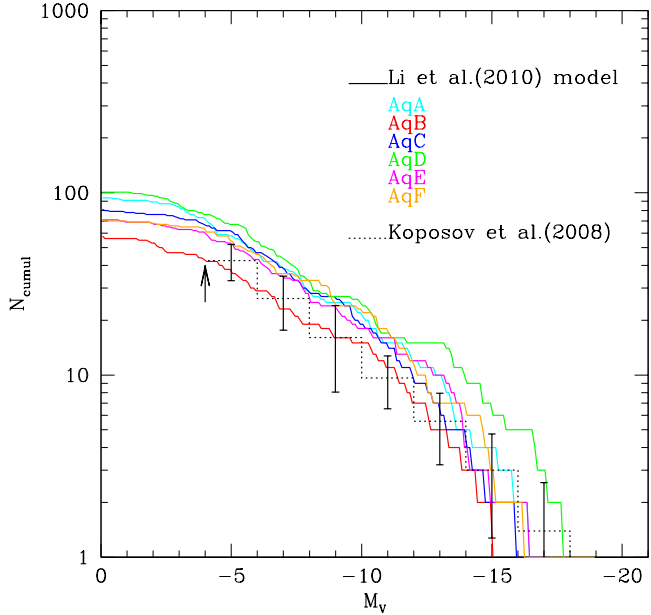}
\caption{\label{fig:lf_li}{Luminosity functions for the six Aquarius
halos in the \citet{li10} model.}}
\end{figure}

\begin{figure}
\includegraphics[width=9.cm] {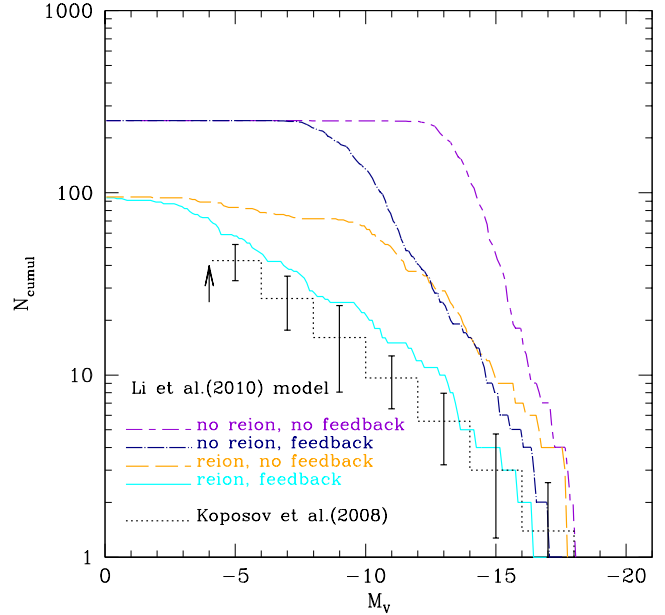}
\caption{\label{fig:lf_reion_fdbk_li}{Luminosity function for the
Aq-A halo in the \citet{li10} model with global reionization and
feedback switched on and off. }}
\end{figure}

It is instructive to test the robustness of our model by comparing
with the results of an independent semi-analytic model implemented in
the same six Aquarius halos. This is the ``Munich model'' described by
\citet{delucia07} and applied to galactic satellites by
\cite{li10}. The \citet{delucia07} model reproduces a variety of
observational data in the local Universe and at high redshift and, as
\citet{delucia08} have shown, with some modifications in the
treatment of disc instabilities and star formation, it reproduces many
observed physical properties of our own Galaxy, including the age and
metallicity distribution of stars in its different components. (The
modifications of \citet{delucia08} do not alter the agreement with
observations shown in \citet{delucia07}). Additional updates of the
reionization and feedback prescriptions were introduced by
\citet{li10} to provide a better match to the observed properties of
Milky Way satellites. A brief description of the relevant aspects of
this model is as follows:

- The cooling model used by \citet{li10} is that originally proposed
by \citet{white91} and used in subsequent versions of the `Munich
model'. Specifically, the cooling rate depends on the temperature and
metallicity of the hot gas, and is regulated by the infall rate in the
`rapid accretion regime'. A comparison between this cooling model and
that adopted by \citet{bower06} is discussed in
\citet{delucia10}. Cooling via molecular hydrogen is not included,
under the assumption that H$_2$ is efficiently photodissociated.

- The reionization is modelled using the methodology of
\citet{croton06}, who adopted the \citet{gnedin00} formalism. In the
\citet{li10} model, however, the reionization is assumed to start at
redshift $z_0 = 15$ and end at $z_r=11.5$, that is, earlier than
implied by the calculation of \citet{gnedin00}, who obtained $z_0=8$
and $z_r=7$.  The reionization scheme assumed in \citet{li10} does not
take into account the local photoionization. However, with the final
choice of parameters, this model becomes similar to our preferred
\galform\, model, fbk:sat/rei:G+L, which includes both global and
local reionization and which reionizes completely at $z \approx10$.

 - Supernova feedback is modelled as in \cite{delucia04}. With the 
adopted parameters, galaxies with virial velocity $V_{vir} <87$ km s$^{-1}$ 
have more heated gas mass per unit stellar mass than in the standard 
\citet{croton06} and \cite{delucia07} models. Their feedback becomes 
inefficient for low mass galaxies, similar to the situation in our 
fbk:sat model (compare the blue dot-long dashed lines in 
Fig.~\ref{fig:lf_reion_fdbk_li} and in the right panel of 
Fig.~\ref{fig:lf_reion_fdbk}).

\citet{li10} applied this 'ejection' model to
a series of high (but much lower than in Aquarius) resolution
simulations of Milky Way-type halos. They found a good match to a
variety of observations, including the satellite luminosity function,
the luminosity-metallicity relation, the radial and size
distributions, and the central mass - luminosity relation.  The
\cite{li10} model yields somewhat higher metallicities than \galform
\, for systems with $L_V \ga 10^{6} L_\odot$ (but is 
able to reproduce the metallicities of the ultra-faint satellites).

We apply here the \citet{li10} model to the six Aquarius halos
analyzed in this study. Fig.~\ref{fig:lf_li} shows the satellite
luminosity functions, constructed in the same way as for \galform\,
(see details in {Section~\ref{sec:results}). Within the scatter there is a 
reasonably good agreement with the observational estimates of
\citet{koposov08}.

Fig.~\ref{fig:lf_reion_fdbk_li} shows the role of supernova feedback
and reionization separately and is similar to Fig.~\ref{fig:lf_reion_fdbk}
in the case of the \galform\, models. There are several common
characteristics between this model and our fiducial model \galform\,
model, fbk:sat/rei:G+L: firstly, feedback from supernovae is more
effective at suppressing the formation of the more luminous systems,
while reionization acts more effectively at the faint end ($M_{\rm V}
\ge -10$); secondly, the reionization alone does not suppress galaxy
formation sufficiently for the model to match the observed luminosity
function. (This is reminiscent of the conclusions of
\citet{benson02b} who also found that global reionization implemented
using the \citet{gnedin00} formalism has a relatively mild effect).

The comparison between different semi-analytical models highlights the 
existence of degeneracies in the way in which different physical processes --
supernova feedback and reionization, in this case -- affect the
properties of the resulting galaxies. However, as we have shown, it is
often possible to break these degeneracies by comparing the model
predictions to a variety of observables rather than to a single
property such as the satellite luminosity function.

\section{Discussion and Conclusions}
\label{sec:concl}

We have implemented a detailed treatment of reionization in the Durham
semi-analytic model of galaxy formation, \galform, and applied it to
the 6 high-resolution simulations of galactic halos of the Aquarius
project. The UV flux produced by the galaxy population is calculated
in a self-consistent way and the contribution from quasars is taken
from the observationally inferred spectrum of \citet{haardt96}. The UV
flux inhibits star formation by: (i)~preventing gas accretion onto low
mass halos, and (ii)~offsetting the cooling rate of the gas already
inside halos.  These effects influence how galaxies form and evolve
and this, in turn, affects the strength of the future UV background,
resulting in a self-consistent calculation of the coupled properties
of the galaxy population and the IGM. 

We use the formalism of
\citet{okamoto08} to calculate the suppression of gas accretion due to
photoheating, and the \cloudy\, software to calculate the net cooling
rate of the gas (i.e., cooling plus heating rates) inside halos.  We
allow for an additional {\it local} UV flux (i.e., over and above that
of the metagalactic UV background) generated by the progenitors of the
Milky Way.  This results in an earlier effective redshift of
reionization for the Milky Way and leads to further suppression of
star formation in satellites, particularly in very low mass systems.
However, in order to reionize the Universe sufficiently early, our
model requires an escape fraction of UV photons of nearly 100\%. This
is higher than current observational estimates. Although these are
uncertain, it seems likely that the high efficiency of photoionization
required by our models may reflect either the limitations of the
inevitable approximations we have made to calculate the reionization
process or the neglect of other processes such as supernova-driven
cosmic ray pressure which
\cite{wade10} argue could also play a role in suppressing galaxy
formation in small halos.

We find that a model with supernova feedback as parameterized by
\citet{bower06} predicts dwarf metallicities that are too low compared to 
observations. This can be rectified by assuming a lower efficiency of
supernova feedback in small systems. We have found that a model with a
saturation in the supernova feedback efficiency ($\beta$ = const) in
dwarf satellite systems and with both local and global reionization
provides the best match to the local luminosity function, the 
metallicity $-$ luminosity relation, and the radial distribution
and central densities of present day dwarf satellites. As it retains
the default feedback efficiency for more massive systems, this model
also matches the large-scale luminosity function.

In the fiducial model the suppression of faint satellites ($M_{\rm V}
> -10$) is achieved by a combination of supernova feedback and
(global+local) photoheating. However, the relative contribution of
these processes varies across the range of dwarf galaxy sizes:
supernova feedback is the dominant process for suppressing the more
massive systems, photoheating plays the main role for the ultra-faint
dwarfs.

While the role of reionization in suppressing galaxy formation in
small halos has long been recognized, the importance of inhomogeneous
reionization has only recently been highlighted
\citep{busha10,munoz09}.  
The local photoheating from Milky Way progenitors is a crucial
ingredient in our model -- unlike in most previous studies of the
formation of satellites in the CDM cosmology, we find that global
reionization by itself does not provide an acceptable solution to the
`missing satellite problem'. While the \citet{bower06} model gives a
good match to the local luminosity function with global reionization
only (see Fig. \ref{fig:lf_B06}), it does not match the $Z -M_{\rm V}$
relation and produces too few dwarf satellites when local sources of
reionization are taken into account. In contrast, the saturated
feedback model matches all of these datasets well. Global reionization
occurs at $z\approx7.8$ and locally at $z\approx 10$.

In Appendix~A, we show that reionization -- with and without local
photoheating -- can be approximated using the standard $v_{\rm
cut}-z_{\rm cut}$ rule \citep{benson02b}, but with parameters that
differ from those typically adopted in other semi-analytical
codes. For example, in the fbk:sat/rei:G model, the best fit is given
by the set ($v_{\rm cut}=34$ km s$^{-1}$, $z_{\rm cut}=7.8$), and in
the fbk:sat/rei:G+L model by the set ($v_{\rm cut}=34$ km s$^{-1}$,
$z_{\rm cut}=10$). The $z_{\rm cut}$ values inferred this way are in
very good agreement with the values derived from the full reionization
treatment (see Fig.~\ref{fig:IGM_fbksat}) and much less
computationally intensive.

A more detailed comparison between our semi-analytical model and other
models can be found in Section~\ref{sec:li} and Appendix~B.  The model of
\citet{li10} is the closest in methodology to the 
\galform\ model presented here and produces similar results.  

The recent renewal of interest in the formation of satellite galaxies
seems to be leading to a consensus that it is possible to reproduce
basic properties of the galaxy population given
\emph{plausible} prescriptions for modelling reionization and its 
effects on galaxy formation
(\citealt{busha10,koposov09,maccio10,busha10,li10}). 
\cite{boylan-kolchin11}, however, have recently presented dynamical
evidence that the largest subhalos in $\Lambda$CDM N-body simulations
are much too concentrated to be able to host the brightest satellites
of the Milky Way. It is unclear at present whether this discrepancy
could be due to statistics (\citealt{parry11}; dynamical data exist
only for a handful of satellites in just one galaxy, the Milky Way) or
whether it points to a more fundamental problem such as the existence
of feedback processes not included in current models or even to a
different nature for the dark matter \citep{lovel11}.

In this work, we have asked the general question of whether, in the
context of the $\Lambda$CDM cosmology, it is possible to account for
the observed properties of satellites not just {\em plausibly} but
specifically within a broad model that has been developed to
understand the galaxy population as a whole, not just the
satellites. This is a more challenging question than merely
constructing a model restricted to satellites, but one that provides a
stronger test of our understanding of galaxy formation. We find, in
agreement with earlier works, that a sufficiently high redshift of
reionization, in combination with feedback from supernovae, can indeed
reduce the number of satellites to a level which agrees with the
data. The specificity of our model allows us to explore additional
properties of satellites, such as their metallicity. We find that the
metallicity-luminosity relation of satellites can only be explained at
the same time as their luminosity function if supernova feedback
saturates at low halo masses. Interestingly, our result raises the
possibility that both the supernova feedback and reionization
(global/local) are scale-dependent processes, which may have important
consequences for our understanding of galaxy formation.

Like other recent models \citep{okamoto09,li10}, ours also provides
an explanation for the apparent common mass scale of dwarf
galaxies. In agreement with the conclusions of
\cite{stringer10}, we find that the weak dependence of $M_{300}$ on
$L$ is a result of a weak dependence of $M_{300}$ on $M_{vir}$. 
Our model predicts that there should, in fact, be a weak trend of increasing 
$M_{300}$ with $L$ and significant scatter in the relation at low 
luminosities. These predictions should be testable once improved 
measurements of $M_{300}$ are available for larger samples of galaxies.

There have been significant and rapid advances over the past five
years in both observations of dwarf satellites and our theoretical
understanding of how they form. Despite this rapid progress, the
conclusions of the previous generation of models remain essentially
correct. The overall results seem independent of the details of any
specific implementation of galaxy formation physics and lead to the
conclusion that the broad visible properties of the satellite
population are reproducible within the current cold dark matter
paradigm.  This represents an important success.  Furthermore, the
cold dark matter cosmogony makes clear predictions for the dark matter
content of dwarf satellites that are mostly independent of the
baryonic physics and are eminently testable by observations
\citep{strigari10,boylan-kolchin11}. An important challenge that
remains for galaxy formation theory is to verify if the good agreement
with dwarf satellite properties can be retained while simultaneously
matching the broader properties of much more massive galaxies and
galaxies at higher redshifts (see, for example, \citet{guo11a}). Given
the complexity of galaxy formation it is only by confronting models
with such a broad range of data that a convincing theory of galaxy
formation will emerge.

\section*{Acknowledgements}
 
We thank Gerry Gilmore for providing us the dwarf satellites
metallicity data in \citet{norris10}. ASF was supported by an STFC
Fellowship at the Institute for Computational Cosmology in Durham and
by a Royal Society Dorothy Hodgkin fellowship at the University of
Cambridge. AJB acknowledges the support of the Gordon and Betty Moore
Foundation. CSF acknowledges a Royal Society Wolfson Research Merit
Award and ERC Advanced Investigator grant COSMIWAY. APC acknowledges
an STFC studentship. GDL acknowledges financial support from the
European Research Council under the European Community's Seventh
Framework Programme (FP7/2007-2013)/ERC grant agreement n. 202781. AH
acknowledges funding support from the European Research Council under
ERC-StG grant GALACTICA-240271. Y-SL was supported by the Netherlands
Organization for Scientific Research (NWO) STARE program
643.200.501. This work was supported in part by an STFC rolling grant
to the Institute for Computational Cosmology of Durham University.

\section*{Appendix A: The $(v_{\rm cut},z_{\rm cut})$ approximation}
\label{sec:AppA}

\begin{figure}
\includegraphics[width=9.cm] {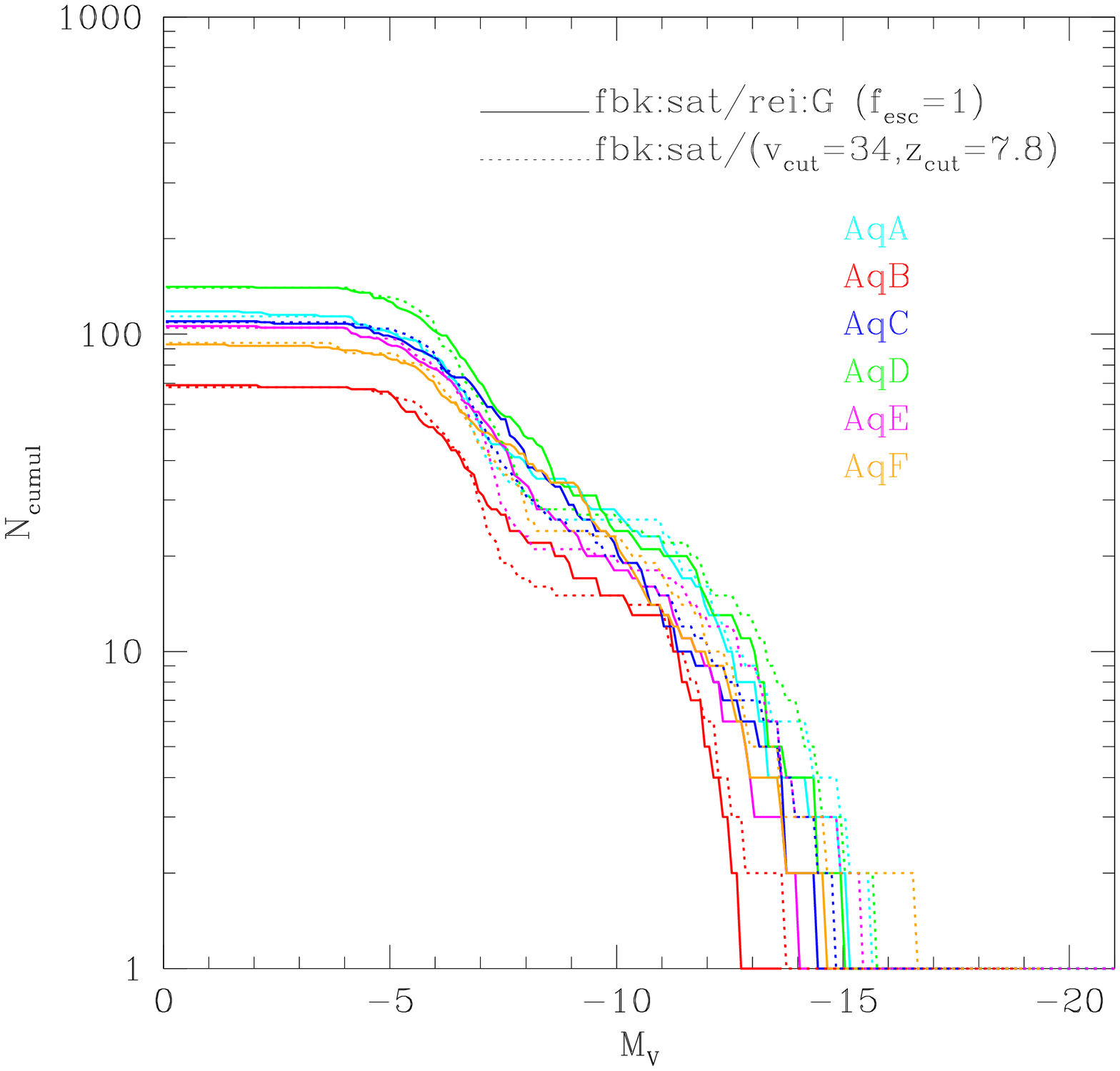}\\
\includegraphics[width=9.cm]{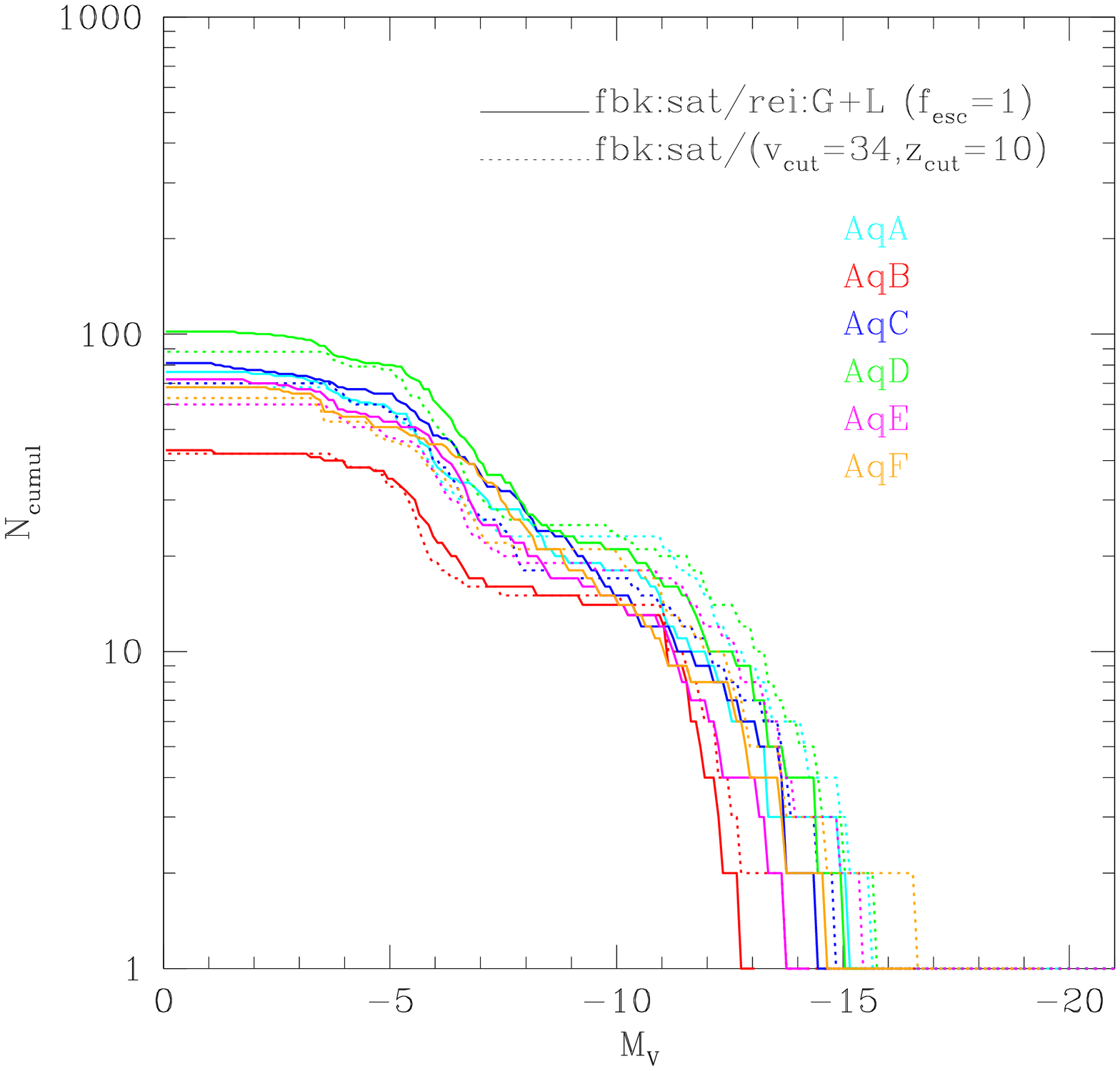}
\caption{\label{fig:lf_vzcut}{Comparison between the models with a
detailed calculation of reionization and the $(v_{\rm cut},z_{\rm
cut})$ approximation. Full lines represent fbk:sat/rei:G models in the top
panel and fbk:sat/rei:G+L models in the bottom panel, respectively. The various
colours correspond to the six Aquarius halos. Dotted lines in both top
and bottom panels represent the models with the $(v_{\rm cut},z_{\rm
cut})$ approximation. The values for this pair of parameters are:
$(v_{\rm cut}, z_{\rm cut})$=(34 km s$^{-1}$, 7.8) in the top panel and
$(v_{\rm cut}, z_{\rm cut})$=(34 km s$^{-1}$, 10) in the bottom
panel.}}
\end{figure}

A self-consistent calculation of reionization in GALFORM was described by
\citet{benson02a}, who showed that this calculation is well
approximated by the ($v_{\rm cut}$, $z_{\rm cut}$) model used
subsequently in the GALFORM code. In this approximation, the baryon
fraction accreted by halos is given by:

\begin{equation}
 f_{\rm b} = \left\{ \begin{array}{ll} 0 & \hbox{if } z < z_{\rm cut}
\hbox{ and } v_{\rm vir} < v_{\rm cut}, \\ \langle f_{\rm b} \rangle &
\hbox{otherwise}, \end{array} \right. 
\end{equation}
where $\langle f_{\rm b} \rangle$ is the mean cosmic baryon
fraction. This `shortcut' greatly speeds up the 
semi-analytic code. It has been adopted in several other 
semi-analytic models such as those by 
\cite{bullock00} and \cite{somerville02}. It is therefore interesting
to check whether the new treatment of reionization implemented in this
paper is still well approximated by a step function of the form above,
and if so, to determine the best fit values of the ($v_{\rm cut},
z_{\rm cut}$) parameters.

We find that the simple ($v_{\rm cut}, z_{\rm cut}$) model still
provides a good approximation to the full calculation of reionization
even when local sources as included. We illustrate this with the fbk:sat
model, although this conclusion is equally valid for other feedback
schemes, including the default power-law feedback scheme, fbk:B06.

Fig.~\ref{fig:lf_vzcut} shows the $(v_{\rm cut},z_{\rm cut})$ approximation 
for the models with saturated feedback and self-consistent global reionization, 
with or without local reionization, fbk:sat/rei:G and fbk:sat/rei:G+L, 
assuming  an escape fraction of ionizing photons of $100\%$. The luminosity 
functions are shown with different colours, each one corresponding to one 
of the six Aquarius halos.

We find that the parameter values $(v_{\rm cut}, z_{\rm cut})=$ (34 km
s$^{-1}$, 7.8) provide a good match to the model with global emission
only, fbk:sat/rei:G.  The best fit $v_{\rm cut}$ is 
slightly higher than the critical circular velocity of 25 km~s$^{-1}$ 
obtained for the Okamoto, Gao, \& Theuns (2008) accretion model. 
This is because in \galform\ photoheating not only prevents accretion 
onto halos below some mass but it also offsets radiative cooling 
losses (and therefore star formation rates) for systems that are massive 
enough to accrete baryons). Note that $z_{cut}$ gives a very good 
approximation to the redshift of reionization calculated self-consistently 
(see Fig. ~\ref{fig:IGM_fbksat}). 

Adding the ionizing flux from the local sources turns out to be equivalent 
to shifting the redshift of reionization to earlier values, while keeping 
$v_{\rm cut}$ roughly constant. The best $(v_{\rm cut}, z_{\rm cut})$ 
approximation for the fbk:sat/rei:G+L case is
$(v_{\rm cut}, z_{\rm cut})=$ (34 km s$^{-1}$, 10).

We find that similar good $(v_{\rm cut}, z_{\rm cut})$ approximations exist 
for the other cases when $f_{\rm esc}$ is varied. Perhaps more surprisingly, 
we find that the $(v_{\rm cut},z_{\rm cut})$ approximation works well for 
a variety of other feedback schemes (including the default fbk:B06 and 
power-laws with varying slopes), regardless or whether or not a local 
photoionizing flux is included, and irrespective of the variety of merger 
histories represented in the Aquarius halos. 
It is also surprising that the approximation works quite well 
down to the regime of the ultra-faint dwarfs, even though it was first 
derived in the regime of classical dwarfs. The success of this approximation
could not have been predicted {\it a priori}, given the differences
between our current models and those of the original \citet{benson02a}
model where the approximation was introduced. 

\section*{Appendix B: Comparison With Other Models}
\label{sec:AppB}

Several other studies of Milky Way satellite galaxies in the context
of the cold dark matter cosmogony have recently been carried out. The
broad consensus from these investigations is that theoretical models
can achieve agreement with the observed distribution of satellite
galaxy luminosities given reasonable assumptions about the process of
star formation and the ability of the reionization of the Universe to
suppress galaxy formation in low mass halos. These studies have
included various pieces of physics thought to be relevant to the
formation of low mass galaxies, but none have performed as detailed
and physically complete calculations as described in this work. Below
we briefly outline the methods and results of these recent works and
contrast them with our own.

\subsection{Busha et al. (2010)}

\cite{busha10} explored the effects of inhomogeneous reionization on
the population of Milky Way satellites using a combination of a
high-resolution N-body simulation of the Milky Way halo and a lower
resolution, larger volume simulation of the universe to assess spatial
variations in the epoch of reionization. They find that reionization
typically occurs between $z=6$ and $z=12$ for Milky Way-sized
halos. To model the formation of galaxies within their dark matter
only simulation, \cite{busha10} assume that halos must have reached a
critical mass (corresponding to the atomic cooling limit of
approximately $10^4$K) prior to reionization. They then assign
luminosities to galaxies using one of two methods. In the first, the
assume a one-to-one mapping between luminosity and peak halo velocity,
$v_{\rm max}$. In the second, more physically motivated approach, a
star formation rate proportional to a power of the halo mass is
assigned to each halo between the time it reaches the critical mass
and the epoch of reionization. The luminosity is found by applying a
stellar population synthesis model and integrating over this star
formation rate. 

\cite{busha10} find that, with a suitable choice of star formation rate
normalization, the second model produces a good match to the faint
end of the observed satellite luminosity function. (The bright end is
underpredicted, but this is understandable as they do not allow for
star formation in massive halos after reionization.) Additionally,
they find that their model produces a good match to the radial
distribution of satellites while simultaneously producing values of
$M_{300}$ in reasonable agreement with the data (although somewhat too
high for the more luminous halos).

In comparison to our treatment, the model of \cite{busha10} is much
more simplistic. However, their results agree with ours at a 
qualitative level -- sufficiently early reionization can reduce the
number of satellites to a level compatible with
observations. Additionally, their conclusion that the abundance of
satellites is very sensitive to the epoch of reionization is
consistent with our findings. For this reason, \cite{busha10}
emphasize the importance of considering inhomogeneous reionization,
consistent with our finding that localized ionization from progenitors
of the Milky Way plays a key role in setting the local reionization
epoch. \cite{busha10} use the Via Lactea II simulation which has
a halo mass of $2.6 \times 10^{12} M_\odot$, somewhat more massive
than the halos considered here, but do not address the issue of
whether their model produces a Milky Way galaxy with the correct
properties (such as stellar mass).

\subsection{Mu\~noz et al. (2009)}

\cite{munoz09} adopt a slightly more involved approach, also utilizing
the Via Lactea II N-body simulation. They consider four channels of
star formation. In low mass halos at early times they allow stars to
form via molecular hydrogen cooling. This process is stopped at $z=20$
when molecular hydrogen is assumed to be dissociated. Prior to
reionization at $z \sim 11$, stars are allowed to form in halos above
the atomic cooling limit, while after reionization, star formation is
restricted to higher mass halos. At late times, some further star
formation resulting from metal line cooling is allowed. In each case,
a simple parameterization is employed in which the mass of stars
formed is proportional to the mass of the halo during the relevant
epoch. Stellar population synthesis models are then used to infer
galaxy luminosities.

With a suitable choice of parameters, \cite{munoz09} find excellent
agreement with the observed satellite luminosity function and show
that molecular hydrogen cooling is important for producing the correct
abundance of low luminosity satellites. They additionally explore the
behaviour of $M_{300}$ as a function of luminosity and find broadly
good agreement with the data although with significantly more scatter
than observed and a trend for $M_{300}$ to be too high in bright
galaxies and too low in faint galaxies.

The work of \cite{munoz09} contains many of the features of our own
work (e.g., $H_{2}$ cooling and the effect of photons generated
locally). Our model, however, includes a rigorous treatment of the
variety of physical process involved in galaxy formation such as gas
cooling, star formation, feedback and global reionization which have
been tested against observations of the galaxy population as a whole
at different redshifts. It is unclear whether the {\em ad hoc}
prescriptions of \cite{munoz09} would lead to realistic galaxies
beyond the satellites of the Milky Way.  

\subsection{Koposov et al. (2009)}

The study by \cite{koposov09} differs from other studies in that it
employs a semi-analytic method to follow the growth and evolution of
the subhalo population inside a halo of final mass
$10^{12}M_\odot$. As we discussed above, it is possible that this mass
may be too small for the Milky Way halo. However, like most other
studies, \cite{koposov09} do not explore whether their model produces
a central galaxy with a stellar mass comparable to that of the Milky
Way. They explore a variety of prescriptions for assigning stars to
subhalos, including simple models in which a constant fraction of
baryons turns into stars and models in which stars can only form in
halos above some critical characteristic velocity after the epoch of
reionization at $z=11$ (either with a sharp transition from star
forming to non-star forming at this critical velocity, or a smoother
transition motivated by the work of \citet{gnedin00}).

They find that the faint end of the luminosity function is made up of
galaxies which formed their stars before the epoch of reionization,
with brighter satellites forming in more massive halos after
reionization. In fact, in their models with a sharp transition at the
critical velocity they see a bimodal luminosity function made up of
these two types of galaxy. In models with a smooth transition the
bimodality is lost and good agreement with the observed luminosity
function is obtained. Their results for $M_{300}$ are consistent with
those of other works (although with less scatter -- a consequence of
their neglect of scatter in the mass-concentration relation, as they
note). 

The approach of \cite{koposov09} is significantly more
phenomenological than that described in this work, but it helps to
confirm, as do other works, that an early epoch of reionization can
plausibly suppress dwarf galaxy formation sufficiently to achieve
agreement with the observations.

\subsection{Macci\`o et al. (2010)}

\cite{maccio10} compare results from three different
semi-analytic models of galaxy formation applied to high-resolution
N-body simulations. Their halos span masses from $1.2$ to $3.6
\times 10^{12}M_\odot$ with a median of $1.7 \times 10^{12}M_\odot$,
and so they could potentially suffer from the same shortcoming as the
Aquarius halos of being of somewhat too low mass. Their semi-analytic
models have been previously matched to the properties of the broader
population of galaxies, but unfortunately, they do not specify
whether or not they produce the correct mass of the central galaxy in
these halos.

In contrast to our approach, the subhalo information is not used to
determine the evolution of satellite galaxies (e.g. to determine
merging timescales). To add a suitable reionization-induced
suppression of galaxy formation, the
\cite{gnedin00} filtering mass prescription is added to each model,
with a reionization history taken from \cite{kravtsov04}. They find
that all three models can achieve a reasonable match to the observed
satellite luminosity function with a reionization epoch of
$z=7.5$. However, they note that the original filtering mass
prescription overestimates the suppressing effects of
reionization. Adopting the currently favoured suppression (which
becomes effective in halos with characteristic velocities below
$\sim30$ km s$^{-1}$), they find that a higher redshift, $z=11$, of
reionization is required to restore a good match to the
data. \cite{maccio10} explore the roles of various physical
ingredients in their models in achieving this match. In particular,
and in agreement with this work, they find that the inclusion of
supernova feedback is crucially important -- without it far too many
luminous galaxies are formed.
 
The work of \cite{maccio10} is closest to our own in terms of the
range of physics modelled and the detail of the treatment. However,
these models lack the potentially important effects of molecular
hydrogen cooling and do not include of a self-consistently computed
reionization history.

\subsection{Guo et al. (2010)}

\cite{guo11a} use a semi-analytic model of galaxy formation to study the
properties of galaxies in dark matter halos spanning a wide range of scales by
utilizing dark matter halo merger trees from the Millennium and 
Millennium-II N-body simulations. Subhalo information is taken 
from the simulations and used to track the merging of satellite galaxies. 
Of relevance to this work, their Millennium-II merger trees resolve 
halos down to a mass of approximately $2\times 10^8M_\odot$, which is 
sufficient to just resolve Milky Way dwarf
satellites. To study Milky Way analogues, \cite{guo11a} select halos from the
Millennium-II simulation which contain a disk-dominated (judged in 
terms of stellar mass) central galaxy with a total stellar mass in the range
$4\times10^{10}M_\odot < M_\star <6\times10^{10}M_\odot$ which results in a
median halo mass of $1.44\times 10^{12}M_\odot$. This once again raises the
potential issue of the halo masses being too small, although in this case if
further observational evidence suggests a higher mass for the Milky Way halo
it would necessitate a recalibration of the \cite{guo11a} galaxy formation
model to reduce the stellar mass of galaxies in dark matter halos of given
mass. The galaxy formation models used are based on models previously used to
successfully model various aspects of the galaxy population. Importantly,
\cite{guo11a} explore how their model performs not only for the Milky Way
satellite population, but also for the broader population of galaxies. They
demonstrate that their model provides a good match to field and cluster galaxy
luminosity functions while simultaneously matching that of Milky Way
satellites.

The key physics of reionization is incorporated into the model of \cite{guo11a}
in by use of the \cite{gnedin00} filtering mass prescription, with a
filtering mass as a function of redshift extracted from the simulations of
\cite{okamoto08}. \cite{guo11a} find that they obtain good agreement with the
luminosity function of Milky Way satellites using their standard reionization
prescription correctly predicts the number of bright satellites, but is
marginally inconsistent with the number of faint satellites (producing
somewhat too many, although the model remains plausible given current
uncertainties in the observational sample). They find that removing the
effects of reionization only affects the abundance of fainter galaxies.
Although \cite{guo11a} do not discuss this point, this implies that supernovae
feedback plays a major role in inhibiting the formation of satellite galaxies,
in agreement with our own findings.

\end{document}